\documentclass[pra,aps,superbib,citeautoscript,preprint]{revtex4-1}

\usepackage{amsmath,bm}
\usepackage{amstext}
\usepackage{epsfig}
\usepackage{xcolor}
\usepackage{subfig}
\usepackage{graphicx}
\usepackage{multirow}
\usepackage{array}
\definecolor{dgreen}{rgb}{0,.5,0}
\definecolor{dred}{rgb}{.7,.0,.0}
\newcommand{\Eq}[1]{Eq.~(\ref{#1})}

\newcommand{\Sec}[1]{Sec. \ref{#1}}

\newcommand{\subi}[2]{#1_{\rm #2}}
\newcommand{\supi}[2]{#1^{\rm #2}}
\newcommand{\subsupi}[3]{#1_{\rm #2}^{\rm #3}}

\begin{document}

\title{
Alternative separation of exchange and correlation energies in range-separated density-functional perturbation theory
}

\author{Yann Cornaton$^1$, Alexandrina Stoyanova$^{1,2}$, 
Hans J\o rgen Aa. Jensen$^3$ and Emmanuel Fromager$^1$
}

\affiliation{\it 
~\\
$^1$Laboratoire de Chimie Quantique,\\
Institut de Chimie, CNRS / Universit\'{e} de Strasbourg,\\
1 rue Blaise Pascal, F-67000 Strasbourg, France\\
$^{2}$Max-Planck-Institut f\"ur Physik komplexer Systeme, \\
                  N\"othnitzer Strasse 38, 01187 Dresden, Germany \\  
$^3$Department of Physics, Chemistry and Pharmacy,\\
University of Southern Denmark,\\
Campusvej 55, DK-5230 Odense M, Denmark
}


\begin{abstract}

An alternative separation of short-range exchange and correlation energies
is used in the framework of second-order range-separated density-functional
perturbation theory.
This alternative separation was initially proposed by
Toulouse {\it et al.} [Theor. Chem.
Acc. {\bf 114}, 305 (2005)] and relies on a long-range interacting
wavefunction instead of the non-interacting Kohn-Sham one.
When second-order corrections to the density are
neglected, the energy expression reduces to a range-separated
double-hybrid (RSDH) type of functional, RSDHf,
where "f" stands for
"full-range integrals" as the regular full-range interaction appears
explicitly in the energy expression when expanded in
perturbation theory. In contrast to usual RSDH functionals, RSDHf
describes the coupling between long- and short-range correlations as an
orbital-dependent contribution. Calculations on the first four noble-gas
dimers show that this coupling has a significant effect on the potential
energy curves in the equilibrium region, improving
the accuracy of binding energies and equilibrium bond distances when
second-order perturbation theory is appropriate.

\end{abstract}

\maketitle


\section{Introduction}\label{intro}

The combination of post-Hartree-Fock (post-HF) methods with density-functional
theory (DFT) by means of range separation has been explored in recent years in
order to improve the long-range part of standard exchange-correlation
functionals. Long-range {\it second-order M\o ller-Plesset}
(MP2)~\cite{Angyan2005PRA,Fromager2008PRA,pra_MBPTn-srdft_janos,Kullie2012CP}, {\it
second-order N-electron valence state perturbation theory}
(NEVPT2)~\cite{nevpt2srdft}, {\it Coupled-Cluster} (CC)~\cite{Goll2005PCCP} as well as
several long-range variants of the {\it random-phase approximation}
(RPA)~\cite{rpa-srdft_scuseria,PRL_rpa-srdft_toulouse,PRA10_Julien_rpa-srDFT}
have thus been
merged with short-range local and semi-local
density-functionals~\cite{Toulouse2004IJQC,Toulouse2004PRA,Goll2005PCCP}, and 
successfully applied to weakly interacting molecular systems~\cite{JCP07_Ian_mp2-srdft_calib_Rg1-Rg2,JCP10_Wuming_rpa-srdft_weak_int,Goll2005PCCP,janosjctc,Mehboob_metal-metal-srDFT}. Even
though such schemes require more computational efforts than standard DFT
calculations, they still keep some of the advantages of the latter in terms of basis
set convergence and basis set superposition error (BSSE).\\

In all the range-separated models mentioned previously the complementary
short-range density-functional describes not only the purely short-range
exchange-correlation energy but also the coupling between long- and
short-range correlations~\cite{Toulouse2004IJQC}. Indeed, as 
the exact
short-range exchange energy is obtained from the non-interacting Kohn-Sham (KS)
determinant, like in standard DFT, the complementary short-range
correlation energy is defined as the difference between the regular
full-range correlation density-functional energy and the purely
long-range one. Local density approximations (LDA) to the
complementary short-range correlation functional can thus be obtained along the
same lines as for the regular correlation energy,
simply by modeling a uniform electron gas with long-range interactions
only~\cite{Toulouse2004IJQC}. Even though
range-separated DFT methods for example
can describe dispersion forces in noble-gas and alkaline-earth-metal dimers,
those based on MP2 or RPA 
with the long-range HF exchange
response kernel (RPAx) often underbind, while a long-range CC
treatment can in
some cases overbind~\cite{PRA10_Julien_rpa-srDFT}. In order to improve on the description of
weak interactions in the equilibrium region, we propose in this work to
describe within MP2, not only the purely long-range correlation but also its
coupling with the short-range one, while preserving a DFT description of
the short-range correlation energy.
This can be achieved rigorously when using an alternative separation of
the exact short-range exchange and correlation energies that relies on the
long-range interacting wave function instead of the KS determinant, as initially proposed
by Toulouse {\it et al.}~\cite{Toulouse2005TCA}. The paper is organized
as follows: in Sec.~\ref{theo} we first motivate the description of the
coupling between long- and short-range correlations within MP2 and then briefly present the usual combination of
long-range MP2 with short-range DFT, leading thus to the definition of
range-separated double hybrid (RSDH) functionals. A new perturbation
expansion of the energy is then derived through second order when using
the alternative  short-range exchange-correlation
energy decomposition of 
Ref.~\cite{Toulouse2005TCA}. Comparison is then made with conventional
double hybrids. The calculation of the orbitals is also discussed.
Following the computational details in Sec.~\ref{comp-details_sec}, results obtained for
the noble-gas dimers within the short-range local density approximation
of Paziani {\it et al.}~\cite{Paziani2006PRB} are presented. Conclusions
are given in Sec.~\ref{Results}.

\section{Theory}\label{theo}

\subsection{Range separation of the second-order M\o ller-Plesset
correlation energy}\label{subsec:RSMP2}

The {\it exact} ground-state energy of an electronic system can be expressed as follows 
\begin{eqnarray}\label{wft-eq}
E 
&=&{\displaystyle \underset{\Psi}{\rm min}\left\{ \langle \Psi\vert
\hat{T}+\hat{W}_{\rm ee}+\hat{V}_{\rm ne}\vert\Psi\rangle
\right\}  
},
\end{eqnarray}
where $\hat{T}$ is the kinetic energy operator,
$\subi{\hat{V}}{ne}$ denotes the nuclear potential operator,
and $\subsupi{\hat{W}}{\rm ee}{}$ is the regular electron-electron
interaction. For systems that are not strongly multi-configurational,
the exact ground-state wave function that minimizes the energy
in Eq.~(\ref{wft-eq}) is reasonably well approximated by the HF determinant
$\Phi_0$ which is obtained when restricting the minimization in
Eq.~(\ref{wft-eq}) to single determinant wavefunctions. Correlation
effects can then be described, for example, in MP perturbation theory where the
first-order correction to the wavefunction contains double excitations
only, which can be expressed in second-quantized form as~\cite{mp_pinkbook}  
\begin{eqnarray}\label{mp1wf}
\vert\Psi^{(1)}\rangle=\frac{1}{2}\sum_{ij,ab}t^{ab(1)}_{ij}\hat{E}_{ai}\hat{E}_{bj}\vert\Phi_0\rangle,
\end{eqnarray}
where
$\hat{E}_{ai}=\hat{a}^\dagger_{a,\alpha}\hat{a}_{i,\alpha}+\hat{a}^\dagger_{a,\beta}\hat{a}_{i,\beta}$
is a singlet excitation operator while $i,j$ and $a,b$ denote occupied
and unoccupied HF orbitals. The MP1 amplitudes are expressed in terms of
the HF orbital energies and the two-electron integrals
$\langle ab\vert ij\rangle=\int\int
{\rm d}\mathbf{r_1}{\rm d}\mathbf{r_2}\phi_a(\mathbf{r_1})\phi_b(\mathbf{r_2})\frac{1}{r_{12}}\phi_i(\mathbf{r_1})\phi_j(\mathbf{r_2})$
as follows
\begin{eqnarray}\label{mp1tabij}
t^{ab(1)}_{ij}=\frac{\langle ab\vert
ij\rangle}{\varepsilon_i+\varepsilon_j-\varepsilon_a-\varepsilon_b},
\end{eqnarray}
and the MP2 correlation energy equals~\cite{mp_pinkbook}
\begin{eqnarray}\label{mp2enertabij}
E^{(2)}&=&\langle\Phi_0\vert\hat{W}_{\rm ee}\vert\Psi^{(1)}\rangle
=
\sum_{ij,ab}
V^{ij}_{ab}
t^{ab(1)}_{ij},
\end{eqnarray}
where the two-electron contributions that are contracted 
with the MP1 amplitudes are  
expressed as 
\begin{eqnarray}\label{Vabij_exp}
V^{ij}_{ab}=2\langle ab\vert
ij\rangle-\langle ab\vert
ji\rangle.
\end{eqnarray}
While the long-range part of the correlation energy can, in many cases, be described
reasonably well within MP2, the accurate description of short-range correlation effects
usually requires the use of CC methods instead of MP2 which
increases the computational cost significantly and requires the use of
large atomic basis sets. On the other hand, standard DFT methods enable a
rather accurate calculation of the short-range
correlation, with a relatively low
computational cost, but they fail in describing long-range correlation
effects. For that reason, Savin~\cite{Savin1996Book} proposed to
separate the two-electron repulsion into
short- and long-range parts 
\begin{eqnarray}\label{ee=lr+sr}
\frac{1}{r_{12}}=w^{\rm lr,\mu}_{\rm ee}(r_{12})+w^{\rm sr,\mu}_{\rm
ee}(r_{12}),
\end{eqnarray}
where $\mu$ is a parameter that controls the range separation, 
so that, for example, a long-range MP2 calculation can be combined
rigorously with a short-range DFT one. The commonly used long-range
interaction~\cite{Toulouse2004PRA} ${w}_{\rm ee}^{\rm lr,\mu}(r_{12})=
{\rm erf}(\mu r_{12})/r_{12}$, which is considered in this work, is
based on the error function but the formalism presented here will be
valid for any separation of the two-electron repulsion.
While the range separation of the
exchange energy is unambiguous, as the latter is linear in the
two-electron interaction, the assignment of long- and short-range correlation effects
to MP2 and DFT, respectively, is less obvious. Indeed, as both MP1 amplitudes
and integrals can be range-separated, according to Eqs.~(\ref{mp1tabij}),
(\ref{Vabij_exp}) and (\ref{ee=lr+sr}),  
\begin{eqnarray}\label{Vandt=lr+sr}
t^{ab(1)}_{ij}&=&
\Big(t^{ab(1)}_{ij}\Big)^{\rm
lr,\mu}
+\Big(t^{ab(1)}_{ij}\Big)^{\rm sr,\mu}
,\nonumber\\
V^{ij}_{ab}&=&
\Big(V^{ij}_{ab}\Big)^{\rm
lr,\mu}+\Big(V^{ij}_{ab}\Big)^{\rm sr,\mu},
\end{eqnarray}
the MP2 correlation energy, that is quadratic in the interaction,
contains purely long-range and purely short-range
contributions as well as 
long-/short-range coupling terms:  
\begin{eqnarray}\label{RSmp2enertabij_lrsr+srlr}\begin{array} {l}
\displaystyle
E^{(2)}=
\sum_{ij,ab}
\Bigg[
\Big(V^{ij}_{ab}\Big)^{\rm lr,\mu}
\Big(t^{ab(1)}_{ij}\Big)^{\rm lr,\mu}
+
\Big(V^{ij}_{ab}\Big)^{\rm sr,\mu}
\Big(t^{ab(1)}_{ij}\Big)^{\rm sr,\mu}\\
\\
\hspace{1.1cm}+
\Big(V^{ij}_{ab}\Big)^{\rm lr,\mu}
\Big(t^{ab(1)}_{ij}\Big)^{\rm sr,\mu}
+
\Big(V^{ij}_{ab}\Big)^{\rm sr,\mu}
\Big(t^{ab(1)}_{ij}\Big)^{\rm lr,\mu}\Bigg].
\end{array}
\end{eqnarray}
Since the last two summations on the right-hand side of
Eq.~(\ref{RSmp2enertabij_lrsr+srlr}) are equal,
\begin{eqnarray}\label{lrsr=srlr}
&&\sum_{ij,ab}
\Big(V^{ij}_{ab}\Big)^{\rm lr,\mu}
\Big(t^{ab(1)}_{ij}\Big)^{\rm sr,\mu}
\nonumber
\\
&&=
\sum_{ij,ab}
\frac{
\big[2\langle ab\vert
ij\rangle^{\rm lr,\mu}-\langle ab\vert
ji\rangle^{\rm lr,\mu}\big]
\langle ab\vert
ij\rangle^{\rm
sr,\mu}}{\varepsilon_i+\varepsilon_j-\varepsilon_a-\varepsilon_b}
\nonumber
\\
&&=
\sum_{ij,ab}
\frac{
2\langle ab\vert
ij\rangle^{\rm sr,\mu}
\langle ab\vert
ij\rangle^{\rm
lr,\mu}}{\varepsilon_i+\varepsilon_j-\varepsilon_a-\varepsilon_b}
\nonumber
\\
&&-\sum_{ij,ab}
\frac{
\langle ab\vert
ij\rangle^{\rm lr,\mu}
\langle ab\vert
ji\rangle^{\rm
sr,\mu}}{\varepsilon_j+\varepsilon_i-\varepsilon_a-\varepsilon_b}
\nonumber
\\
&&=
\sum_{ij,ab}
\Big(V^{ij}_{ab}\Big)^{\rm sr,\mu}
\Big(t^{ab(1)}_{ij}\Big)^{\rm lr,\mu},
\end{eqnarray}
the range-separated MP2 correlation energy can finally be rewritten as 
\begin{eqnarray}\label{RSmp2enertabij}
E^{(2)}=
\sum_{ij,ab}&&
\Big(V^{ij}_{ab}\Big)^{\rm lr,\mu}
\Big(t^{ab(1)}_{ij}\Big)^{\rm lr,\mu}\nonumber\\
&&+
\Big(V^{ij}_{ab}\Big)^{\rm sr,\mu}
\Big(t^{ab(1)}_{ij}\Big)^{\rm sr,\mu}\nonumber\\
&&+2
\Big(V^{ij}_{ab}\Big)^{\rm sr,\mu}
\Big(t^{ab(1)}_{ij}\Big)^{\rm lr,\mu}.
\end{eqnarray}
In conventional range-separated density-functional perturbation
theory~\cite{Angyan2005PRA,Fromager2008PRA,pra_MBPTn-srdft_janos},
the long-range correlation energy only is
described within MP2 while the short-range correlation
and its coupling with the long-range one are modeled by a complementary local or
semi-local density-functional. While it is important, in terms of computational cost, to describe the purely short-range
correlation energy within DFT, the coupling term could in principle be treated
within MP2. In the particular case of a van der Waals dimer like Ar$_2$,
for example, this term is not expected to contribute significantly to
the dispersion interaction energy at long distance as the short-range integral contributions will
vanish. However, in the equilibrium region ($R_e\approx7.1
a_0$~\cite{Tang2003JCP}),
the average correlation distance between the valence electrons located on different Ar
atoms is approximately $R_e-2R_a\approx 4.4a_0$, where $R_a=1.34a_0$ is
the atomic radius of Ar. The former distance should then be
compared with the inverse $1/\mu$ of the range separation parameter that
defines qualitatively what are long and short-range
interactions~\cite{Pollet:2002p2223}. In
conventional range-separated calculations
$\mu\approx0.4-0.5a_0^{-1}$~\cite{dft-Fromager-JCP2007a,
nancycalib,Kullie2012CP} which
leads to $1/\mu\approx2.0-2.5a_0$. As a result, 
short-range integrals will be of the same order of magnitude or smaller
than long-range ones so that, in this
case, they can contribute significantly to the
dispersion interaction energy. It is thus relevant to raise the question
whether an MP2
description of the long-/short-range correlation coupling is not
preferable to a DFT one. Note that, as readily seen from Eq.~(\ref{RSmp2enertabij}), this would only require the computation of
the short-range integrals that would then be contracted with the
long-range MP1 amplitudes. In this respect, such a new scheme would
still be based on a long-range MP2 calculation so that the advantages of
the conventional range-separated MP2-DFT model~\cite{Angyan2005PRA}
relative to regular MP2,
like a faster convergence with respect to the basis set and a smaller
BSSE, would be preserved. After a short introduction to conventional
range-separated density-functional perturbation theory in
Sec.~\ref{srDFT}, we will show in Secs.~\ref{alternativesrHxcdecomp_sec}
and \ref{rsdhf_sec} how the long-/short-range MP2 coupling term can be rigorously introduced
into the energy expansion through second order by means of a different separation of the exact
short-range exchange and correlation energies.   

\subsection{Range-separated density-functional perturbation theory}\label{srDFT}

In conventional range-separated DFT~\cite{Savin1996Book}, which we refer
to as short-range DFT (srDFT), the {\it exact} ground-state energy is rewritten as
\begin{eqnarray}\label{srDFT-eq}
E 
&=&{\displaystyle \underset{\Psi}{\rm min}\left\{ \langle \Psi\vert
\hat{T}+\hat{W}^{\rm lr,\mu}_{\rm ee}+\hat{V}_{\rm ne}\vert\Psi\rangle
+E^{\rm sr,\mu}_{\rm Hxc}[n_{\Psi}]\right\}  
}\nonumber\\
&=&{\displaystyle \langle \Psi^\mu\vert
\hat{T}+\hat{W}^{\rm lr,\mu}_{\rm ee}+\hat{V}_{\rm ne}\vert\Psi^\mu\rangle
+E^{\rm sr,\mu}_{\rm Hxc}[n_{\Psi^\mu}],  
}
\end{eqnarray}
where $\subsupi{\hat{W}}{\rm ee}{lr,\mu}$ is the long-range electron-electron
interaction operator. The complementary short-range
Hartree-exchange-correlation (srHxc) density-functional energy is
denoted $\subsupi{E}{Hxc}{sr,\mu}[n]$. While Eq.~(\ref{wft-eq}) is
recovered in the $\mu\rightarrow+\infty$ limit, the other limit $\mu=0$
corresponds to regular KS-DFT as the long-range interaction vanishes and
the srHxc functional reduces to the conventional Hxc one. 
As a zeroth-order approximation, the minimization in Eq.~(\ref{srDFT-eq}) can
be performed over single determinant wave functions, leading to the
HF-srDFT scheme (referred to as RSH for {\it range-separated
hybrid} in Ref.~\cite{Angyan2005PRA}). The minimizing HF-srDFT determinant $\Phi_0^\mu$
fulfills the following long-range HF-type equation 
\begin{eqnarray}\label{0thord-SCeq}\begin{array}{l}
\hat{H}_0|\Phi_0^\mu\rangle=\mathcal{E}_0|\Phi_0^\mu\rangle,
\\
\\
\hat{H}_0=\hat{T}+\subsupi{\hat{U}}{HF}{lr,\mu}+\subi{\hat{V}}{ne}+
\displaystyle
\int {\rm d}{\bf
r}\,
\dfrac{\delta \subsupi{E}{Hxc}{sr,\mu}}{\delta n({\bf
r})}
[n_{\Phi_0^\mu}]
\,\hat{n}({\bf
r}),
\end{array}
\end{eqnarray}
where $\subsupi{\hat{U}}{HF}{lr,\mu}=\sum_{pq}
\sum_i\big(2\langle pi \vert qi\rangle^{\rm lr,\mu}
-\langle pi \vert
iq\rangle^{\rm lr,\mu}
\big)\hat{E}_{pq}$ is the long-range
analog of the non-local HF potential operator, constructed with the
occupied
HF-srDFT orbitals, 
and $\hat{n}({\bf r})$ denotes the density
operator.
The long-range dynamical correlation effects, which are not
described at the HF-srDFT level, can then be treated within a long-range
MP-type perturbation theory~\cite{Goll2005PCCP,Angyan2005PRA}.
For that purpose, we 
introduce a perturbation strength $\alpha$ and define the auxiliary
energy~\cite{Angyan2005PRA}
\begin{eqnarray}\label{Ealphamu_var-pt}
E^{\alpha,\mu} = \underset{\Psi}{\rm min}\Big\{ 
&&
\langle \Psi\vert
\hat{T}+
(1-\alpha)\subsupi{\hat{U}}{HF}{lr,\mu}
+
\alpha\subsupi{\hat{W}}{ee}{lr,\mu}
\vert\Psi\rangle
\nonumber\\
&&+\langle\Psi\vert\hat{V}_{\rm ne}
\vert\Psi\rangle
+E^{\rm sr,\mu}_{\rm Hxc}[n_{\Psi}]\Big\}.  
\end{eqnarray}
The minimizing wave function
$\Psi^{\alpha,\mu}$ in Eq.~(\ref{Ealphamu_var-pt}) can be obtained
self-consistently from the following non-linear eigenvalue-type equation 
\begin{eqnarray}\label{srSCeq_with_alpha}\begin{array} {l}
\Bigg(\hat{T}+(1-\alpha)\subsupi{\hat{U}}{HF}{lr,\mu}+
\alpha\subsupi{\hat{W}}{ee}{lr,\mu}+\hat{V}_{\rm ne}\\
\\
+\displaystyle
\int {\rm d}{\bf
r}\,
\dfrac{\delta \subsupi{E}{Hxc}{sr,\mu}}{\delta n({\bf
r})}
[n_{\Psi^{\alpha,\mu}}]
\,\hat{n}({\bf
r})
\Bigg)\vert\Psi^{\alpha,\mu}\rangle=\mathcal{E}^{\alpha,\mu}\vert\Psi^{\alpha,\mu}\rangle.
\end{array}
\end{eqnarray}
It is readily seen, from Eqs.~(\ref{0thord-SCeq}) and
(\ref{srSCeq_with_alpha}), that in the $\alpha=0$ limit,
$\Psi^{\alpha,\mu}$ reduces to the HF-srDFT determinant $\Phi_0^\mu$, while,
according to Eqs.~(\ref{srDFT-eq}) and (\ref{Ealphamu_var-pt}), the auxiliary energy becomes, for $\alpha=1$, the {\it exact}
ground-state energy and $\Psi^{\alpha,\mu}$ reduces to the minimizing
wave function $\Psi^{\mu}$ in Eq.~(\ref{srDFT-eq}). Using the intermediate normalization condition
\begin{eqnarray}\label{intnorma_alpha}
\langle
\Phi_0^\mu\vert\Psi^{\alpha,\mu}\rangle=1,\hspace{0.5cm} 0\leq
\alpha\leq 1,
\end{eqnarray}
it was shown~\cite{Angyan2005PRA,Fromager2008PRA,pra_MBPTn-srdft_janos,Fromager2011JCP} that the wave
function can be expanded through second order as follows
\begin{eqnarray}\label{wf-pertalpha}
|\Psi^{\alpha,\mu}\rangle&=&|\Phi_0^\mu\rangle+\alpha|\Psi^{\rm
(1)lr,\mu}\rangle+\alpha^2|\Psi^{\rm
(2)\mu}\rangle+\mathcal{O}(\alpha^3),
\end{eqnarray}
where the first-order contribution is the long-range analog of the
MP1 wavefunction correction
\begin{eqnarray}\label{lrmp1wf}
\vert\Psi^{(1)\rm lr,\mu}\rangle=\frac{1}{2}\sum_{ij,ab}
\Big(t^{ab(1)}_{ij}\Big)^{\rm
lr,\mu}
\hat{E}_{ai}\hat{E}_{bj}\vert\Phi^\mu_0\rangle,
\end{eqnarray}
that is computed with HF-srDFT orbitals and orbital energies. Indeed, 
according to the Brillouin theorem, the density remains unchanged through first
 order, leading
to the following Taylor expansion, through second order, for the
density: 
\begin{eqnarray}\label{density-pert}
n_{\Psi^{\alpha,\mu}}({\bf r})&=&n_{\Phi^\mu_0}({\bf r})+\alpha^2\delta n^{(2)\mu}({\bf
r})+\mathcal{O}(\alpha^3),
\end{eqnarray}
so that self-consistency effects in Eq.~(\ref{srSCeq_with_alpha}) do
not contribute to the wave function through first order. Non-zero
contributions actually appear through second order in the wave
function~\cite{Fromager2011JCP}. Finally, the auxiliary energy can be
expanded as~\cite{Angyan2005PRA,Fromager2008PRA}
\begin{eqnarray}\label{ptalpha_second}\begin{array}{l}
\displaystyle 
{E}^{\alpha,\mu}
={E}^{(0)\mu}+\alpha{E}^{(1)\mu}+\alpha^2{E}^{(2)\mu}+\mathcal{O}(\alpha^3),
\end{array}
\end{eqnarray}
where, when considering the $\alpha=1$ limit, the HF-srDFT energy is recovered through first order 
\begin{eqnarray}\label{HF-srDFT-eq2}
\subsupi{E}{\tiny HF}{\tiny srDFT}&=&{E}^{(0)\mu}+{E}^{(1)\mu}\\
&=&\langle\Phi_0^\mu|\hat{T}+\subsupi{\hat{W}}{ee}{lr,\mu}+\subi{\hat{V}}{ne}
|\Phi_0^\mu\rangle
+\subsupi{E}{Hxc}{sr,\mu}[n_{\Phi_0^\mu}]\nonumber,
\end{eqnarray}
and the second-order correction to the energy is the purely long-range MP2 correlation energy
\begin{eqnarray}\label{secondorderenermp}
E^{(2)\mu}=
\sum_{ij,ab}&&
\Big(V^{ij}_{ab}\Big)^{\rm lr,\mu}
\Big(t^{ab(1)}_{ij}\Big)^{\rm lr,\mu}
.
\end{eqnarray}
In summary, the HF-srDFT scheme defines an approximate one-parameter RSH
exchange-correlation energy which combines exact long-range exchange
with complementary srDFT exchange-correlation
energies:
\begin{eqnarray}\label{xcener_RSH}
\subsupi{E}{xc,
RSH}{\mu}&=&
-\sum_{ij}\langle ij \vert
ji\rangle^{\rm lr,\mu}
+\subsupi{E}{x}{sr,\mu}[n_{\Phi_0^\mu}]\nonumber\\
&&+\subsupi{E}{c}{sr,\mu}[n_{\Phi_0^\mu}]. 
\end{eqnarray}
Including second-order terms leads to the MP2-srDFT energy
expression~\cite{Fromager2008PRA}
(referred to as RSH+MP2 in Ref.~\cite{Angyan2005PRA}),
\begin{eqnarray}\label{MP2-srDFT-ener}
\subsupi{E}{\tiny MP2}{\tiny srDFT}&=&\subsupi{E}{\tiny HF}{\tiny
srDFT}+E^{(2)\mu},
\end{eqnarray}
which defines, according to Eq.~(\ref{secondorderenermp}), an
approximate one-parameter range-separated double hybrid exchange-correlation energy expression
\begin{eqnarray}\label{xcener_RSDH}
\subsupi{E}{xc,
RSDH}{\mu}&=&
-\sum_{ij}\langle ij \vert ji\rangle^{\rm lr,\mu}
+
\sum_{ij,ab}
\Big(V^{ij}_{ab}\Big)^{\rm lr,\mu}
\Big(t^{ab(1)}_{ij}\Big)^{\rm lr,\mu}\nonumber\\
&&+\subsupi{E}{x}{sr,\mu}[n_{\Phi_0^\mu}]+\subsupi{E}{c}{sr,\mu}[n_{\Phi_0^\mu}],
\end{eqnarray}
where both the purely short-range correlation energy and its coupling
with the long-range one are described by the complementary short-range
correlation density-functional.
\subsection{Alternative decomposition of the short-range
exchange-correlation energy}\label{alternativesrHxcdecomp_sec}

Since the exact short-range exchange-correlation density-functional is
unknown, local and semi-local approximations have been developed in
order to perform practical srDFT
calculations~\cite{Savin1996Book,Toulouse2004IJQC,Heyd2003JCP,Heyd2004JCP,Toulouse2004PRA,Gori-Giori2006PRA,Goll2005PCCP,Goll2009JCP}.
The former are based on the following decomposition of the srHxc energy
\begin{eqnarray} \label{srDFTfunhxcdef}
\subsupi{E}{Hxc}{sr,\mu}[n]&=&\subsupi{E}{H}{sr,\mu}[n]+\subsupi{E}{x}{sr,\mu}[n]+\subsupi{E}{c}{sr,\mu}[n],\\
\subsupi{E}{H}{sr,\mu}[n]&=&\frac{1}{2}\int\int {\rm d}{\mathbf{r}}\,{\rm d}{\mathbf{r'}}\,n(\mathbf{r})\,n(\mathbf{r'})\,\subsupi{w}{ee}{sr,\mu}\left(|{\bf r}-{\bf r'}|\right),\nonumber\\
\subsupi{E}{x}{sr,\mu}[n]&=&\langle\supi{\Phi}{KS}[n]|\subsupi{\hat{W}}{ee}{sr,\mu}|\supi{\Phi}{KS}[n]\rangle-\subsupi{E}{H}{sr,\mu}[n],\nonumber
\end{eqnarray}
where the short-range correlation energy is defined with respect to the
KS determinant $\Phi^{\rm KS}[n]$ like in standard DFT. As an
alternative, Toulouse, Gori-Giorgi and
Savin~\cite{Toulouse2005TCA,Gori-Giorgi2009IJQC} proposed a
decomposition which relies on the ground state $\Psi^{\rm \mu}[n]$ of the long-range interacting system whose density equals $n$: 
\begin{eqnarray}\label{EsrHxc_decomps2}
\subsupi{E}{Hxc}{sr,\mu}[n]&=&
\langle\Psi^{\rm \mu}[n]|\subsupi{\hat{W}}{ee}{sr,\mu}|\Psi^{\rm \mu}[n]\rangle
+\subsupi{E}{\rm c, md}{sr,\mu}[n]. 
\end{eqnarray}

The first term in the right-hand side of Eq.~(\ref{EsrHxc_decomps2}) was
referred to as "multideterminantal" ("md") short-range exact
exchange~\cite{Gori-Giorgi2009IJQC}. As
shown in the following, it contains not only the short-range exchange
energy but also coupling terms between long-
and short-range correlations. Note that, according to
Eq.~(\ref{EsrHxc_decomps2}), the complementary "md" short-range
correlation functional differs from the conventional one defined in \Eq{srDFTfunhxcdef}: 
\begin{eqnarray}\label{Esrc_md}
\subsupi{E}{\rm c, md}{sr,\mu}[n]&=&\subsupi{E}{\rm c}{sr,\mu}[n]+\langle\Phi^{\rm KS}[n]|\subsupi{\hat{W}}{ee}{sr,\mu}|\Phi^{\rm KS}[n] \rangle \nonumber \\
&&-\langle\Psi^{\rm \mu}[n]|\subsupi{\hat{W}}{ee}{sr,\mu}|\Psi^{\rm \mu}[n]\rangle.
\end{eqnarray}
This expression has been used in Refs.~\cite{Toulouse2005TCA,Paziani2006PRB} for developing  
short-range "md" LDA correlation functionals. Returning to the exact energy expression in \Eq{srDFT-eq}, the srHxc energy can be written as
\begin{eqnarray}\label{EsrHxc}
\subsupi{E}{Hxc}{sr,\mu}[n_{\Psi^\mu}]&=&
\langle\Psi^{\rm \mu}|\subsupi{\hat{W}}{ee}{sr,\mu}|\Psi^{\rm
\mu}\rangle+\subsupi{E}{c,md}{sr,\mu}[n_{\Psi^\mu}],
\end{eqnarray}
using the decomposition in \Eq{EsrHxc_decomps2} and the first
Hohenberg--Kohn theorem~\cite{Hohenberg1964PR} which ensures, according
to Eq.~(\ref{srSCeq_with_alpha}) in the $\alpha=1$ limit, that
$\Psi^\mu$ is the ground state of a long-range interacting system and
therefore
\begin{eqnarray}\label{lrhkth}
\Psi^\mu[n_{\Psi^\mu}]&=&\Psi^\mu.
\end{eqnarray}
When adding long- and short-range interactions in
\Eq{srDFT-eq}, the {\it exact} ground-state energy expression becomes~\cite{Toulouse2005TCA} 
\begin{eqnarray}\label{rsH}
E&=&\langle\Psi^\mu|\hat{T}+\subi{\hat{W}}{ee}+\subi{\hat{V}}{ne}
|\Psi^\mu\rangle+\subsupi{E}{c,md}{sr,\mu}[n_{\Psi^\mu}].
\end{eqnarray}
As shown in Sec.~\ref{rsdhf_sec}, using such an expression, in combination
with the MP2-srDFT perturbation expansion of the wave function in
Eq.~(\ref{wf-pertalpha}),
will enable us to define a new class of RSDH functionals where both
the purely long-range MP2 correlation energy and the long-/short-range
MP2 coupling term appear
explicitly in the energy expansion through second order.
Let us finally mention that, as shown by Sharkas {\it et
al.}~\cite{2blehybrids_Julien}, regular double hybrid density-functional
energy expressions
can be derived when considering a scaled interaction $\lambda_1/r_{12}$
instead of a long-range one based on the error function. With the notations of
Ref.~\cite{2blehybrids_Fromager2011JCP}
, the short-range
exchange-correlation energy decomposition in
Eq.~(\ref{EsrHxc_decomps2}) becomes, for the scaled interaction,
\begin{equation}\label{mdXCcm1def}
\overline{E}^{\lambda_1}_{\rm Hxc}[n]= 
(1-\lambda_1)\langle \Psi^{\lambda_1}[n]\vert \hat{W}_{\rm ee} 
\vert \Psi^{\lambda_1}[n]\rangle
+\overline{E}^{\lambda_1}_{\rm c,md}[n],
\end{equation}
where $\Psi^{\lambda_1}[n]$ is the ground state of the
$\lambda_1$-interacting system whose density equals $n$. As further
discussed in
Sec.~\ref{connection_regular_2ble_sec}, the new class of RSDH which is
derived in this work can be connected to conventional two-parameter double hybrids by
means of Eq.~(\ref{mdXCcm1def}).
\subsection{New class of range-separated double hybrid density-functionals}\label{rsdhf_sec}
In order to derive a new perturbation expansion of the energy based on Eq.~(\ref{rsH}), we 
introduce a modified auxiliary energy
\begin{eqnarray}\label{newptalpha}\begin{array}{l}
\displaystyle 
\tilde{E}^{\alpha,\mu}
=
{E}^{\alpha,\mu}
-\subsupi{E}{Hxc}{sr,\mu}[n_{\Psi^{\alpha,\mu}}]
\\
\\
\hspace{1.3cm}+\alpha\dfrac{\langle\Psi^{\alpha,\mu}|\subsupi{\hat{W}}{ee}{sr,\mu}|\Psi^{\alpha,\mu}\rangle}{\langle\Psi^{\alpha,\mu}|\Psi^{\alpha,\mu}\rangle}+\subsupi{E}{c,md}{sr,\mu}[n_{\Psi^{\alpha,\mu}}],
\end{array}
\end{eqnarray}
which reduces, like ${E}^{\alpha,\mu}$, to the {\it exact} ground-state energy in the $\alpha=1$
limit. As argued in Sec.~\ref{subsec:RSMP2}, RSDH functionals are well adapted to
the description of weakly interacting systems. We also pointed out that, for
standard $\mu$ values, the short-range integrals
associated to dispersion interactions may be of the same
order of magnitude or smaller than their long-range counterparts. It is
therefore relevant, for such systems, to consider the 
short-range interaction as a first-order contribution in perturbation theory, like the long-range
fluctuation operator in MP2-srDFT (see Eq.~(\ref{Ealphamu_var-pt})). This justifies the multiplication by $\alpha$ of the short-range
interaction expectation value in Eq.~(\ref{newptalpha}). It also ensures
that truncating the Taylor expansion of the modified auxiliary energy
through second order is as relevant as for the auxiliary energy used in
MP2-srDFT. Let us mention that the perturbation theory presented in the
following differs from MP2-srDFT only by the energy expansion. Both
approaches will indeed be based on the same wavefunction perturbation
expansion. As discussed further in Sec.~\ref{subsec:calc_orbitals}, it
is, in principle, possible to correct both the wave function and the
energy consistently when combining {\it optimized effective potential} (OEP)
techniques with range separation. Note also, in
Eq.~(\ref{newptalpha}), the
normalization factor in front of the short-range interaction expectation
value, which must be introduced since the intermediate normalization is
used (see Eq.~(\ref{intnorma_alpha})). From the wavefunction
perturbation expansion in Eq.~(\ref{wf-pertalpha}), we thus obtain
the orthogonality condition
$\langle\Phi^\mu_0\vert\Psi^{\rm(1)lr,\mu}\rangle=0$ and, as a result,
the following Taylor expansion:
\begin{eqnarray}\label{srexpvalue_alpha}\begin{array}{l}
\displaystyle
\frac{\left\langle\Psi^{\alpha,\mu}\right|\subsupi{\hat{W}}{ee}{sr,\mu}\left|\Psi^{\alpha,\mu}\right\rangle}{\langle
\Psi^{\alpha,\mu}\vert\Psi^{\alpha,\mu}\rangle}=\left\langle\Phi^\mu_0\right|\subsupi{\hat{W}}{ee}{sr,\mu}\left|\Phi^\mu_0\right\rangle
\\
\\
+2\alpha\langle\Phi^\mu_0\vert\subsupi{\hat{W}}{ee}{sr,\mu}\vert\Psi^{\rm
(1)lr,\mu}\rangle
+\mathcal{O}(\alpha^2).
\end{array}
\end{eqnarray}
In addition, according to Eq.~(\ref{density-pert}), the short-range
density-functional energy difference can be expanded through second
order as
\begin{eqnarray}\label{ptexp-srfuncdiff_alpha}\begin{array}{l}
\Big(\subsupi{E}{c,md}{sr,\mu}
-\subsupi{E}{Hxc}{sr,\mu}\Big)[n_{\Psi^{\alpha,\mu}}]
=
\Big(\subsupi{E}{c,md}{sr,\mu}-\subsupi{E}{Hxc}{sr,\mu}\Big)[n_{\Phi_0^\mu}]
\\
\\
\displaystyle+\alpha^2\int {\rm d}{\bf r}\,\Bigg(
\dfrac{\delta \subsupi{E}{c,md}{sr,\mu}}{\delta n({\bf r})}
-\dfrac{\delta \subsupi{E}{Hxc}{sr,\mu}}{\delta n({\bf
r})}\Bigg)[n_{\Phi_0^\mu}]
\;\delta
n^{(2)\mu}({\bf r})
\\
\\
+\mathcal{O}(\alpha^3).
\end{array}
\end{eqnarray}
As a result, we obtain from Eqs.~(\ref{ptalpha_second}),
(\ref{newptalpha}), (\ref{srexpvalue_alpha}) and
(\ref{ptexp-srfuncdiff_alpha}) a new Taylor expansion for the energy
\begin{eqnarray}\label{newptalpha_second}\begin{array}{l}
\displaystyle 
\tilde{E}^{\alpha,\mu}
=\tilde{E}^{(0)\mu}+\alpha\tilde{E}^{(1)\mu}+\alpha^2\tilde{E}^{(2)\mu}+\mathcal{O}(\alpha^3),
\end{array}
\end{eqnarray}
where
\begin{eqnarray}\label{newPTexpener_alpha}\begin{array} {l}
\tilde{E}^{(0)\mu}
={E}^{(0)\mu}+
\Big(\subsupi{E}{c,md}{sr,\mu}-\subsupi{E}{Hxc}{sr,\mu}\Big)[n_{\Phi_0^\mu}],
\\
\\
\tilde{E}^{(1)\mu}
={E}^{(1)\mu}+
{\langle\Phi_0^\mu\vert\subsupi{\hat{W}}{ee}{sr,\mu}\vert\Phi_0^\mu\rangle},
\\
\\
\tilde{E}^{(2)\mu}
={E}^{(2)\mu}
+2{\langle\Phi_0^\mu\vert\subsupi{\hat{W}}{ee}{sr,\mu}\vert\Psi^{(1)\rm
lr,\mu}\rangle}
\\
\\
\displaystyle+\int {\rm d}{\bf r}\,\Bigg(
\dfrac{\delta \subsupi{E}{c,md}{sr,\mu}}{\delta n({\bf r})}
-\dfrac{\delta \subsupi{E}{Hxc}{sr,\mu}}{\delta n({\bf
r})}\Bigg)[n_{\Phi_0^\mu}]
\;\delta
n^{(2)\mu}({\bf r})
.\\
\end{array}
\end{eqnarray}
According to Eq.~(\ref{HF-srDFT-eq2}), in the $\alpha=1$ limit, we recover through first order what can be referred to as a
RSH energy expression with {\it full}-range integrals (RSHf)
\begin{eqnarray}\label{rshfener}
{E}_{\rm RSHf}&=&\tilde{E}^{(0)\mu}+\tilde{E}^{(1)\mu}\\
&=&\langle\Phi_0^\mu|\hat{T}+\subi{\hat{W}}{ee}+\subi{\hat{V}}{ne}
|\Phi_0^\mu\rangle+\subsupi{E}{c,md}{sr,\mu}[n_{\Phi_0^\mu}],\nonumber
\end{eqnarray}
which defines the approximate one-parameter RSHf exchange-correlation energy
\begin{eqnarray}\label{xcener_RSHf}
\subsupi{E}{xc,
RSHf}{\mu}&=&
-\sum_{ij}\langle ij \vert
ji\rangle
+\subsupi{E}{c, md}{sr,\mu}[n_{\Phi_0^\mu}]
. 
\end{eqnarray}
Turning to the second-order energy corrections in
Eq.~(\ref{newPTexpener_alpha}), the second term on the right-hand side
of the third equation can be identified, by analogy
with Eq.~(\ref{mp2enertabij}), as the long-/short-range MP2 coupling term that was introduced
in the range-separated expression of the MP2 correlation energy in
Eq.~(\ref{RSmp2enertabij}):
\begin{eqnarray}\label{Elr-srMP2final}
2\langle\Phi^\mu_0|\subsupi{\hat{W}}{ee}{sr,\mu}\vert\supi{\Psi}{(1)lr,\mu}\rangle
&=&
2\sum_{ij,ab}
\Big(V^{ij}_{ab}\Big)^{\rm sr,\mu}
\Big(t^{ab(1)}_{ij}\Big)^{\rm lr,\mu}.
\end{eqnarray}
Note that this coupling is here expressed in terms of the HF-srDFT
orbitals and orbital energies. We thus  
deduce from Eq.~(\ref{secondorderenermp}) the final expression for the second-order correction to the energy:
\begin{eqnarray}\label{newPT2ener}\begin{array} {l}
\displaystyle
\tilde{E}^{(2)\mu}
=
\sum_{ij,ab}
\Bigg[\Big(V^{ij}_{ab}\Big)^{\rm lr,\mu}+2\Big(V^{ij}_{ab}\Big)^{\rm
sr,\mu}\Bigg]
\Big(t^{ab(1)}_{ij}\Big)^{\rm lr,\mu}
\\
\\
\displaystyle+\int {\rm d}{\bf r}\,\Bigg(
\dfrac{\delta \subsupi{E}{c,md}{sr,\mu}}{\delta n({\bf r})}
-\dfrac{\delta \subsupi{E}{Hxc}{sr,\mu}}{\delta n({\bf
r})}\Bigg)[n_{\Phi_0^\mu}]\;\delta n^{(2)\mu}({\bf r}).
\end{array}
\end{eqnarray}
Note that Eqs.~(\ref{rshfener}) and ~(\ref{newPT2ener}) provide an
energy expression which is {\it exact} through second order. Note also
that, in contrast to
MP2-srDFT~\cite{pra_MBPTn-srdft_janos,janos_2nplus1rule_MBPTn-srdft}
for which the auxiliary energy in Eq.~(\ref{Ealphamu_var-pt}) has a variational expression, the $2n+1$ rule is not fulfilled
here as the modified auxiliary energy expression in Eq.~(\ref{newptalpha}) 
is not variational. In other words, the Hellmann--Feynman theorem does not
hold in this context. If we neglect the second-order correction to the density,
that is well justified for molecular systems which are not
multi-configurational~\cite{Fromager2011JCP,Goll2005PCCP}, 
we obtain a RSDH energy expression involving, through the exchange energy, {\it full}-range integrals (RSDHf)
\begin{eqnarray}\label{RSDHfapprox}\begin{array}{l}
{E}_{\rm RSDHf}= {E}_{\rm RSHf}\\
\displaystyle
+
\sum_{ij,ab}
\Bigg[\Big(V^{ij}_{ab}\Big)^{\rm lr,\mu}+2\Big(V^{ij}_{ab}\Big)^{\rm
sr,\mu}\Bigg]
\Big(t^{ab(1)}_{ij}\Big)^{\rm lr,\mu}
.
\end{array}
\end{eqnarray}
Note that, according to Eqs.~(\ref{secondorderenermp}),
(\ref{MP2-srDFT-ener}) and (\ref{RSDHfapprox}), 
the same long-range correlation energy expression, that is based on MP2,
is used in both RSDHf and
MP2-srDFT schemes. The former differs from the latter, in the second-order correlation
energy correction, only by the coupling between long- and
short-range correlations that is now described within MP2, 
\begin{eqnarray}\label{RSDHf-RSDHcomparmp2}
{E}_{\rm RSDHf}-\subsupi{E}{\tiny MP2}{\tiny srDFT}&=& 
{E}_{\rm RSHf}
+2\sum_{ij,ab}
\Big(V^{ij}_{ab}\Big)^{\rm
sr,\mu}
\Big(t^{ab(1)}_{ij}\Big)^{\rm lr,\mu}
\nonumber\\
&&-\subsupi{E}{\tiny HF}{\tiny srDFT}
.
\end{eqnarray}
Hence, according to Eq.~(\ref{xcener_RSHf}), the RSDHf energy expression in Eq.~(\ref{RSDHfapprox}) defines a new type of approximate
one-parameter RSDH exchange-correlation energy, 
\begin{eqnarray}\label{xcener_rsdhf}\begin{array} {l}
E_{\rm xc, RSDHf}^{\mu}
=
\displaystyle
-\sum_{ij}\langle ij \vert
ji\rangle
\\
\displaystyle
+\sum_{ij,ab}
\Bigg[\Big(V^{ij}_{ab}\Big)^{\rm lr,\mu}+2\Big(V^{ij}_{ab}\Big)^{\rm
sr,\mu}\Bigg]
\Big(t^{ab(1)}_{ij}\Big)^{\rm lr,\mu}
\\
\\
+\subsupi{E}{c, md}{sr,\mu}[n_{\Phi_0^\mu}]
,
\end{array}
\end{eqnarray}
which is, in addition, self-interaction free as 
both long- and short-range exchange energies are treated exactly. 
In conclusion, according to Eqs.~(\ref{xcener_RSDH}) and
(\ref{xcener_rsdhf}), RSDHf differ from MP2-srDFT in terms of
exchange and correlation energies as follows: 
\begin{eqnarray}\label{compar_rsdhf_rsdh}
E_{\rm x, RSDHf}^{\mu}-E_{\rm x,
RSDH}^{\mu}
&=&
-\sum_{ij}\langle ij \vert ji\rangle^{\rm sr,\mu}
-\subsupi{E}{x}{sr,\mu}[n_{\Phi_0^\mu}],\nonumber\\
E_{\rm c, RSDHf}^{\mu}-E_{\rm c,
RSDH}^{\mu}
&=&
\subsupi{E}{c, md}{sr,\mu}[n_{\Phi_0^\mu}]\nonumber\\
&&+
2\sum_{ij,ab}
\Big(V^{ij}_{ab}\Big)^{\rm
sr,\mu}
\Big(t^{ab(1)}_{ij}\Big)^{\rm lr,\mu}
\nonumber\\
&&-\subsupi{E}{c}{sr,\mu}[n_{\Phi_0^\mu}].
\end{eqnarray}
It is known that, in practice, standard hybrid functionals only use a
fraction of exact exchange. The situation here is quite different as 
a part of the correlation energy is treated explicitly in perturbation
theory. In this respect, it is not irrelevant to investigate RSDH schemes that use $100\%$ of
full-range exact exchange. Numerical results presented in \Sec{Results} actually support this statement. Nevertheless, as shown in
Appendix~\ref{2rsdhf_appendix}, a two-parameter RSDHf (2RSDHf) model can be
formulated when introducing a fraction $\lambda$ of exact short-range
exchange energy. This leads to the following exchange-correlation
expression 
\begin{eqnarray}\label{xcener_2rsdhf}
E_{\rm xc, 2RSDHf}^{\mu,\lambda}
&=&
-\sum_{ij}\Big[\langle ij \vert ji\rangle^{\rm lr,\mu}
+\lambda
\langle ij \vert ji\rangle^{\rm sr,\mu}\Big]
\nonumber\\
&&
+\sum_{ij,ab}
\Bigg[\Big(V^{ij}_{ab}\Big)^{\rm lr,\mu}+2\lambda\Big(V^{ij}_{ab}\Big)^{\rm
sr,\mu}\Bigg]
\Big(t^{ab(1)}_{ij}\Big)^{\rm lr,\mu}
\nonumber\\
\nonumber\\
&&+(1-\lambda)\Big(\subsupi{{E}}{\rm x}{sr,\mu}[n_{\Phi_0^\mu}]+\subsupi{{E}}{\rm
c}{sr,\mu}[n_{\Phi_0^\mu}]\Big)
\nonumber\\
&&+
\lambda\subsupi{E}{c,md}{sr,\mu}[n_{\Phi_0^\mu}].
\end{eqnarray}
As readily seen in Eq.~(\ref{xcener_2rsdhf}), the 2RSDHf scheme reduces
to MP2-srDFT and RSDHf models in the $\lambda=0$ and $\lambda=1$ limits,
respectively. Note that, for any value of $\lambda$, the long-range
exchange and purely long-range correlation energies are fully treated at the HF and MP2
levels, respectively. As a result, the second parameter $\lambda$ can
only be used for possibly improving, in practice, the description of the
complementary short-range
energy. Such a scheme is not further investigated in this paper 
and is left for future work.

\subsection{Connection with conventional double
hybrids}\label{connection_regular_2ble_sec}

As shown in
Refs.~\cite{2blehybrids_Fromager2011JCP,AC_2blehybrids_Yann},
the exchange-correlation energy
expression that is used in conventional {\it two-parameter double hybrids} (2DH), 
\begin{eqnarray}\label{xcener_likeDH}\begin{array} {l}
\displaystyle
E_{\rm {xc},\mbox{\tiny 2DH}}^{a_{\rm x},a_{\rm c}}= 
-a_{\rm x}
\sum_{ij}\langle ij \vert
ji\rangle
+(1-a_{\rm x})E_{\rm x}[n]
\\
\displaystyle
\hspace{1.6cm}+(1-a_{\rm c})E_{\rm c}[n]+
a_{\rm c}
\sum_{ij,ab}
V^{ij}_{ab}
t^{ab(1)}_{ij},
\end{array}
\end{eqnarray}
can be derived within density-functional perturbation theory when the
following conditions are fulfilled: 
\begin{eqnarray}\label{cond_axac}
0\leq a_{\rm x}&\leq& 1, 
\nonumber\\
0\leq a_{\rm c}&\leq& a^2_{\rm x}.
\end{eqnarray}
This is achieved when applying the MP2 approach to an electronic system whose
interaction is scaled as $\lambda_1/r_{12}$ where 
\begin{eqnarray}\label{lambda1_functionaxac}\begin{array} {l}
\lambda_1=a_{\rm x}- \sqrt{a^2_{\rm x}-a_{\rm c}}.
\end{array}
\end{eqnarray}
As already mentioned in Sec.~\ref{alternativesrHxcdecomp_sec}, an analogy can be made between regular double hybrids and the RSDHf
model derived in Sec.~\ref{rsdhf_sec}, considering first, according to Eq.~(\ref{xcener_rsdhf}), a
fraction of 100\% for the
exact exchange energy
\begin{eqnarray}\label{ax_analog_rsdhf}
a_{\rm x}&=&1,
\end{eqnarray}
which, according to Eq.~(\ref{lambda1_functionaxac}), leads to 
\begin{eqnarray}\label{ac_analog_rsdhf}
a_{\rm c}=1-(1-\lambda_1)^2
.
\end{eqnarray}
When switching from the scaled to the long-range interaction,
\begin{eqnarray}\label{lambda1_eq_lrmu}
\lambda_1/r_{12}\;\rightarrow\; 
w^{\rm lr,\mu}_{\rm ee}(r_{12}),
\end{eqnarray}
or equivalently
\begin{eqnarray}\label{1mlambda1_eq_srmu}
(1-\lambda_1)/r_{12}\;\rightarrow\; 
w^{\rm sr,\mu}_{\rm ee}(r_{12}),
\end{eqnarray}
the fraction of MP2 correlation energy in a regular double hybrid
becomes, according to Eqs.~(\ref{ac_analog_rsdhf}) and
(\ref{1mlambda1_eq_srmu}),
\begin{eqnarray}\label{corr_ener_connection_2blehybrid_mp2}
a_{\rm c}
\sum_{ij,ab}
V^{ij}_{ab}
t^{ab(1)}_{ij}
&\;\rightarrow\;& 
\sum_{ij,ab}
V^{ij}_{ab}
t^{ab(1)}_{ij}\nonumber\\
&&-
\sum_{ij,ab}
\Big(V^{ij}_{ab}\Big)^{\rm sr,\mu}
\Big(t^{ab(1)}_{ij}\Big)^{\rm sr,\mu}
,
\end{eqnarray}
and, from Eq.~(\ref{mdXCcm1def}) as well as   
Eqs.~(10), (13) and (36) in
Ref.~\cite{2blehybrids_Fromager2011JCP}, 
we obtain for the DFT correlation term  
\begin{eqnarray}\label{corr_ener_connection_2blehybrid_dft}
\nonumber\\
(1-a_{\rm c})E_{\rm c}[n]\approx\overline{E}^{\lambda_1}_{\rm c,md}[n] &\;\rightarrow\;&\subsupi{E}{c,
md}{sr,\mu}[n], 
\end{eqnarray}
where, on the left-hand side of
Eq.~(\ref{corr_ener_connection_2blehybrid_dft}), the uniform coordinate
scaling in the density has been neglected, as in conventional
double hybrids~\cite{2blehybrids_Julien,2blehybrids_Fromager2011JCP}. As
a result, using Eqs.~(\ref{mp2enertabij}) and (\ref{RSmp2enertabij}), we recover by simple analogy the RSDHf exchange-correlation energy expression of
Eq.~(\ref{xcener_rsdhf}). 
\subsection{Calculation of the orbitals}\label{subsec:calc_orbitals}

As mentioned in Sec.~\ref{rsdhf_sec}, the RSDHf exchange-correlation energy expression in
Eq.~(\ref{xcener_rsdhf})
would be {\it exact} through second order 
if the second-order corrections to the density had not been neglected.
In practical calculations, further approximations must be considered.
The first one concerns the short-range "md" correlation energy
functional for which local approximations have been
developed~\cite{Toulouse2005TCA,Paziani2006PRB}. The second one is
related to the calculation of the
HF-srDFT orbitals which is "exact" only if Eq.~(\ref{0thord-SCeq}) is
solved with the {\it exact} srHxc density-functional potential. In this
work, an approximate short-range LDA (srLDA) potential will be used. Note that, as an
alternative, OEP techniques could also be applied for obtaining possibly
more accurate srHxc potential and orbitals~\cite{Toulouse2005TCA,Gori-Giorgi2009IJQC}. 
The simplest procedure, referred to as HF-srOEP, would consist in
optimizing the short-range potential at the RSHf
level, in analogy to the {\it density-scaled two-parameter} HF-OEP
(DS2-HF-OEP) scheme of Ref.~\cite{2blehybrids_Fromager2011JCP}, that is,
without including long-range and long-/short-range MP2
contributions to the srOEP. There is
no guarantee that the corresponding MP2-srOEP scheme will perform better than RSDHf,
simply because correlation effects may affect the orbitals and the
orbital energies significantly.
Moreover, the srOEP would also depend on the approximation used
for the short-range "md" correlation functional. Numerical calculations
would be necessary to assess the accuracy of the MP2-srOEP scheme. Work
is currently
in progress in this direction.

\section{Computational details}\label{comp-details_sec}

The RSDHf exchange-correlation energy expression in
Eq.~(\ref{xcener_rsdhf}) has
been implemented in a development version of the DALTON program
package~\cite{daltonpack}. The complementary "md" srLDA functional of
Paziani {\it et al.}~\cite{Paziani2006PRB} has been used. The HF-srLDA
orbitals, that are used in the computation of both RSDHf and MP2-srLDA
energies, were
obtained with the srLDA exchange-correlation functional of Toulouse {\it
et al.}~\cite{Toulouse2004IJQC}. The latter was also used for calculating
the complementary srDFT part of the MP2-srLDA energy. Note that the "md"
srLDA functional is not expected to reduce to the srLDA correlation
functional of Ref.~\cite{Toulouse2004IJQC} in the $\mu=0$ limit, as it should
in the exact theory according to Eq.~(\ref{Esrc_md}). Indeed, while the
former is based on quantum Monte Carlo calculations, the latter was
analytically parametrized from CC calculations performed on a long-range
interacting uniform electron
gas. Interaction energy curves have been
computed for the first four noble-gas dimers. Augmented correlation-consistent polarized quadruple-$\zeta$ basis sets ("aug-cc-pVQZ") of Dunning and co-workers~\cite{Dunning1989JCP,Kendall1992JCP,Woon1993JCP,Woon1994JCP,Koput2002JPCA,Wilson1999JCP} have been used. Comparison is made with regular MP2 and CCSD(T)
approaches. The counterpoise method has been used for removing the BSSE.  
Equilibrium distances ($R_{\rm e}$), equilibrium interaction energies ($D_{\rm e}$) and
harmonic vibrational wavenumbers ($\omega_{\rm e}$) have been obtained 
through fits by a second-order Taylor expansion of the interaction energy 
\begin{eqnarray}\begin{array}{l}
\displaystyle
U(R)=-D_{\rm
e}+\frac{1}{2}k(R-R_{\rm e})^2,\\
\\
\displaystyle
\omega_{\rm e}=\frac{1}{2\pi
c}\sqrt{\frac{k}{\mu_0}},
\end{array}
\end{eqnarray}
where $c$ is the speed of light in the vacuum and $\mu_0$ is the reduced
mass of the dimer. An extended
Levenberg-Marquardt algorithm~\cite{Marquardt2012JMC} on a set of points
from $R_{\rm e}-0.02a_0$ to $R_{\rm e}+0.02a_0$ by steps of $0.01a_0$
has been used.
$C_6$ dispersion coefficients have been calculated by fitting the
expression $U(R)=-C_6/R^6$ with the same algorithm on a set of points from $30.0$ to $60.0a_0$ by steps of $5.0a_0$~\cite{Angyan2005PRA}.
Hard core radii have been calculated by searching for the
distance $\sigma$ at which $U(\sigma)=0$.
The analytical potential curves of Tang and Toennies~\cite{Tang2003JCP}
are used as reference. 

\section{Results and discussion}\label{Results}

\subsection{Choice of the $\mu$ parameter} \label{mu-choice}

In this section we discuss the choice of the range-separation parameter
$\mu$ for practical MP2-srDFT and RSDHf calculations. 
Following the prescription of Fromager \textit{et
al.}~\cite{dft-Fromager-JCP2007a},
which consists in assigning short-range correlation effects to the
density-functional part of the energy to
the maximum possible extent, we investigate, for the first four noble-gas atoms, the variation of the second-order correlation
energy, in both MP2-srDFT and RSDHf, when increasing $\mu$ from zero.
Results are shown in Fig.~\ref{Emu_atcurves}.    
The recipe given in Ref.~\cite{dft-Fromager-JCP2007a} for
the definition of an optimal $\mu$ value consists in determining
the largest $\mu$ value, in systems that do not exhibit long-range
correlation effects,
for which the energy correction induced by the long-range post-HF treatment
remains relatively small (1 $mE_h$ was used as
threshold in Ref.~\cite{dft-Fromager-JCP2007a}). Such a value ensures that the Coulomb hole is essentially
described within DFT. 
For MP2-srDFT, this
criterion leads to the value $\mu=0.4a_0^{-1}$ (see
Fig.~\ref{Emu_atcurves}) which is
in agreement with previous works based on multi-configuration srDFT
calculations~\cite{dft-Fromager-JCP2007a,JCPunivmu2}. 
As shown by Str\o msheim {\it et al.}~\cite{Teale:2011}, this value 
ensures that most of the dispersion in 
He$_2$ is assigned to the long-range
interaction. 
It is
relatively close to $0.5a_0^{-1}$, which is used in RSH+lrMP2 and
RSH+lrRPA
calculations~\cite{Angyan2005PRA,JCP07_Ian_mp2-srdft_calib_Rg1-Rg2,JCP10_Wuming_rpa-srdft_weak_int,JCP11_Julien_ringCC_srDFT,PRA10_Julien_rpa-srDFT} 
and that has been calibrated for reproducing at the exchange-only RSH
level atomization energies
of small molecules~\cite{nancycalib}. In the case of RSDHf, 
the second-order correlation energy deviates from 
zero (within an accuracy of 1 $mE_h$) for much smaller $\mu$ values
(about 0.15$a_0^{-1}$) and is, up to $0.4a_0^{-1}$, completely dominated by the
lr-sr MP2 coupling term. One may thus conclude that the prescription of 
Ref.~\cite{dft-Fromager-JCP2007a} leads to different optimal
$\mu$ values when considering RSDHf energies. It is in fact more
subtle. Let us first note that Fig.~\ref{Emu_atcurves} can be
rationalized when considering the Taylor expansion of the
range-separated MP2 correlation energy as $\mu\rightarrow 0$, 
which
leads to 
\begin{eqnarray}\label{rsmp2correnerexpansion}\begin{array} {l}
\displaystyle
\sum_{ij,ab}
\Big(V^{ij}_{ab}\Big)^{\rm lr,\mu}
\Big(t^{ab(1)}_{ij}\Big)^{\rm lr,\mu}
=A_6\mu^6+\mathcal{O}(\mu^7),\\
\displaystyle
2\sum_{ij,ab}
\Big(V^{ij}_{ab}\Big)^{\rm
sr,\mu}
\Big(t^{ab(1)}_{ij}\Big)^{\rm lr,\mu}
=B_3\mu^3+B_4\mu^4\\
\hspace{4.3cm}+B_5\mu^5+B_6\mu^6+\mathcal{O}(\mu^7),
\end{array}
\end{eqnarray}
where the expansion coefficients are expressed in terms of the KS
orbitals and orbital energies (see
Appendix~\ref{rsmp2_Taylor_mu_appendix}). The fact
that the lr-sr MP2 coupling term varies as $\mu^3$ for small $\mu$
values, while the purely
long-range MP2 correlation energy varies as $\mu^6$, explains the earlier
deviation from zero observed for the RSDHf second-order correlation
energy. It also helps to realize that the RSDHf second-order correlation
energy is not the right
quantity to consider when applying the recipe of Fromager \textit{et
al.}~\cite{dft-Fromager-JCP2007a} with a threshold of 1 $mE_h$, as its variation for small $\mu$
values depends more on the
short-range interaction than the long-range one. Note that the lowest
order terms in Eq.~(\ref{rsmp2correnerexpansion}) simply
arise from the Taylor expansion of the long- and short-range
interactions~\cite{Toulouse2004PRA} 
\begin{eqnarray}\label{taylorexpsrlr_lowest_order}
\displaystyle 
w^{\rm lr,\mu}_{\rm ee}(r_{12})
&=&
\frac{2}{\sqrt{\pi}}\Big(\mu-\frac{1}{3}\mu^3r_{12}^2+\mathcal{O}(\mu^5)\Big),
\nonumber\\
\displaystyle w^{\rm sr,\mu}_{\rm
ee}(r_{12})&=&\frac{1}{r_{12}}+\mathcal{O}(\mu),
\end{eqnarray}
which can be inserted into the range-separated MP2 correlation energy
expression in 
Eq.~(\ref{RSmp2enertabij}). 
In conclusion, the order of magnitude of the purely long-range MP2
correlation energy only 
should be used for choosing $\mu$ with the energy criterion of 
Ref.~\cite{dft-Fromager-JCP2007a}, exactly like in MP2-srDFT. Hence, 
RSDHf calculations presented in the following have been performed with $\mu=0.4a_0^{-1}$.
Note, however, that the lr-sr MP2 curve in Fig.~\ref{Emu_atcurves} is still of interest as it exhibits, for
all the noble gases considered, an inflection point around $0.4-0.5a_0^{-1}$
which can be interpreted as the transition from short-range to long-range
interaction regimes as $\mu$ increases. This confirms that choosing
$\mu=0.4a_0^{-1}$ for the purpose of assigning the Coulomb hole to the
short-range interaction is relevant. A change in curvature was
actually expected as, in both $\mu=0$
and $\mu\rightarrow+\infty$ limits, the coupling term strictly equals zero. 
Let us finally mention that natural orbital occupancies computed through
second order can also be used to
justify the choice of $\mu$ in MP2-srDFT~\cite{dft-Fromager-JCP2007a,Fromager2011JCP}, and consequently in RSDHf, as both
methods rely on the same wavefunction perturbation expansion.   

\subsection{Analysis of the interaction energy curves} \label{Ng-curves}

Interaction energy curves have been computed for the first four noble-gas dimers
with the various range-separated hybrid and double hybrid schemes
presented in Secs.~\ref{srDFT} and \ref{rsdhf_sec}, using the srLDA
approximation. As discussed in Sec.~\ref{mu-choice}, the $\mu$ parameter
was set to $0.4a_0^{-1}$. Comparison is made with conventional MP2 and
CCSD(T) results. The curves are shown in Fig.~\ref{potcurve-comp}. As
expected from Ref.~\cite{Angyan2005PRA}, HF-srLDA
and RSHf models do not describe weak interactions between the two atoms as they both neglect long-range correlation
effects. 
Interestingly RSHf is even more repulsive than HF-srLDA. According to
Eqs.~(\ref{xcener_RSH}) and (\ref{xcener_RSHf}) the two methods differ by (i) the short-range exchange energy,
which is treated with an LDA-type functional within HF-srLDA and exactly within RSHf, and (ii) the complementary short-range correlation energy which
is in both cases treated with an LDA-type functional but includes within
HF-srLDA more correlation than in RSHf. The RSHf curve could thus be expected to get
closer to the regular HF curve which was shown to be more repulsive than
HF-srLDA, at least for Ne and heavier noble-gas
dimers~\cite{Angyan2005PRA}.
When the lr-sr MP2 coupling term is added to the RSHf
energy (RSHf + lr-srMP2 curve in Fig.~\ref{potcurve-comp}), the interaction energy curve
becomes less repulsive than the HF-srLDA one. It means that using an
exact short-range exchange energy in combination with a purely
short-range density-functional correlation energy and a lr-sr MP2 coupling term,
like in RSHf + lr-srMP2, can provide substantially different interaction
energies when comparison is made with a range-separated DFT method where the complementary
short-range exchange-correlation energy is entirely described with an approximate
density-functional, like in HF-srLDA. Let us stress that this
difference, that is
significant for He$_2$, is the one obtained when comparing RSDHf and MP2-srLDA interaction energies
since both are computed with the same purely long-range MP2 term which
enables recovery of the dispersion interactions on the MP2 level (see
Eq.~(\ref{RSDHf-RSDHcomparmp2})).
As RSDHf binds more than MP2-srLDA, RSDHf binding energies
become closer to the experimental values for
He$_2$ and Ne$_2$. In the former case, RSDHf and CCSD(T)
curves are almost on top of
each other and are the closest to experiment. For Ar$_2$ and
Kr$_2$, as MP2-srLDA already performs well, RSDHf slightly overbinds.
The error
on the binding energy is still in absolute value comparable to that obtained
with CCSD(T) while the equilibrium bond distances remain accurate and
closer to experiment than CCSD(T) (see Table~\ref{tabcurvref}). We
should stress here that comparison with CCSD(T) results is not
completely fair as those are not fully basis-set converged while MP2-srLDA and RSDHf
results are almost converged, which is, of course, a nice feature of range-separated
schemes. This point will be discussed further in
Sec.~\ref{subsec_performance_rsdhf}.\\ Following Kullie and
Saue~\cite{Kullie2012CP}, we computed for analysis purposes
long- and short-range exchange-correlation energy contributions to the
RSDHf and MP2-srLDA interaction energies when varying the $\mu$ parameter 
for Ar$_2$ at $R=7.013a_0$ and $10a_0$. In the former case which
corresponds to the RSDHf equilibrium geometry we first notice in
Fig.~\ref{Emu_Ar2curves} (a) that, for $\mu=0.4a_0^{-1}$, the computed srLDA and exact short-range exchange energies are
fortuitously equal and strongly attractive. As a result the difference
between RSDHf and MP2-srLDA interaction energies is only due to the complementary short-range
correlation energy. Since the srLDA correlation energy numerically equals for
$\mu=0.4a_0^{-1}$ the lr-sr MP2 coupling term, again fortuitously, this
difference reduces to the complementary "md" srLDA correlation energy
(see Eq.~(\ref{compar_rsdhf_rsdh})) which
is attractive and thus makes RSDHf bind more than MP2-srLDA. 
It is then
instructive to vary $\mu$ around 0.4$a_0^{-1}$. As $\mu$
decreases the srLDA exchange becomes increasingly attractive
while the srLDA correlation interaction energy does not vary
significantly (see Fig.~\ref{Emu_Ar2curves} (a)). As a result MP2-srLDA, which reduces to standard LDA in
the $\mu=0$ limit, increasingly overbinds when $\mu\leq0.2a_0^{-1}$. On the
contrary, RSDHf is less attractive as $\mu$ decreases from 0.4$a_0^{-1}$ and
becomes repulsive for $\mu\leq0.25a_0^{-1}$. In the latter case, the purely long-range     
MP2 interaction energy becomes negligible, as already observed for Kr$_2$ by Kullie
and Saue~\cite{Kullie2012CP}. The attractive lr-sr MP2 coupling term also decreases
in absolute value but remains significant for $0.1\leq\mu\leq0.2a_0^{-1}$, as
expected from the analysis in Sec.~\ref{mu-choice}. As the attractive "md" srLDA
correlation contribution does not vary significantly as $\mu$ decreases
while the exact short-range exchange becomes less attractive, RSDHf does
become repulsive. Note that, for $\mu=0$, the "md" srLDA correlation
interaction energy does not reduce exactly to the srLDA correlation one, as it
should in the exact theory, simply because the functionals were not
developed from the same uniform electron gas model (see
Sec.~\ref{comp-details_sec}). 
Beyond $\mu=0.4a_0^{-1}$ the lr-sr MP2 coupling
decreases in absolute value as well as the difference between RSDHf and
MP2-srLDA interaction energies which is consistent with the fact that
both methods reduce to standard MP2 in the $\mu\rightarrow+\infty$
limit.\\  
Let us now focus on the lr-sr MP2 contribution to the interaction
energy. According to Eq.~(\ref{Elr-srMP2final}) it is constructed from the product of long-
and short-range integrals associated to dispersion-type double
excitations, namely simultaneous single excitations on each monomer. For
a given bond distance $R$ we can define from the atomic radius $R_a$ a reference $\mu_{\rm
ref}=1/(R-2R_a)$ parameter for which both long- and short-range integrals
are expected to be significant and therefore give the largest
absolute value for the lr-sr MP2 interaction energy. When using
$R_a=1.34a_0$ in the case of Ar, we obtain $\mu_{\rm ref}=0.23a_0^{-1}$ and $0.14a_0^{-1}$
for $R=7.013$ and $10a_0$, respectively. These values are in
relatively good agreement with the minima of the lr-sr MP2 term in
Figs.~\ref{Emu_Ar2curves} (a) and
(b). Even though, for $\mu=0.4a_0^{-1}$, the lr-sr
MP2 coupling term does not reach its largest contribution to the equilibrium
interaction energy, it is still far from
negligible. It actually contributes for about half of the binding energy, which explains why RSDHf and MP2-srLDA curves differ
substantially in the equilibrium region. At the larger $R=10a_0$ distance,
$\mu=0.4a_0^{-1}$ is too large, when compared to $0.14a_0^{-1}$, for the
lr-sr MP2 coupling to contribute significantly (see Fig.~\ref{Emu_Ar2curves} (b)). Similarly the
complementary short-range exchange-correlation terms are relatively
small and completely dominated by
the purely long-range MP2 term. This explains why RSDHf and MP2-srLDA
interaction energy curves get closer as $R$ increases.

\subsection{Performance of the RSDHf model}\label{subsec_performance_rsdhf}

Equilibrium binding energies ($D_{\rm e}$), equilibrium bond distances
($R_{\rm e}$), harmonic vibrational wavenumbers ($\omega_{\rm e}$), hard core radii ($\sigma$) as well as $C_6$ dispersion coefficients have been computed at the RSDHf level for the first four noble-gas
dimers. Results are presented in Table~\ref{tabcurvref} where comparison is made with MP2-srLDA,
MP2 and CCSD(T). As mentioned previously, RSDHf binds more than MP2-srLDA
which is an improvement  
for both He$_2$ and Ne$_2$, in terms of equilibrium bond distances, binding
energies and hard core radii. Interestingly, Toulouse {\it et
al.}~\cite{PRA10_Julien_rpa-srDFT} observed similar trends
when replacing in the MP2-srDFT calculation the long-range MP2 treatment
by a long-range RPA description including the long-range HF exchange
response kernel (RPAx), or when using a long-range CCSD(T) description.
For He$_2$, the RSDHf harmonic vibrational wavenumber is 
closer to experiment than the MP2-srLDA one, but not for Ne$_2$. In this case it is still
more accurate than regular MP2. For Ar$_2$ and Kr$_2$, RSDHf
overestimates both equilibrium binding energies and harmonic vibrational wavenumbers but the
errors are comparable in absolute value to the CCSD(T) ones. On the
other hand, the equilibrium
bond distances remain relatively accurate when compared to MP2-srLDA
values. The two-parameter extension of
RSDHf in Eq.~(\ref{xcener_2rsdhf}) might be a good compromise for improving
MP2-srLDA while avoiding overbinding. Calibration studies on a larger test
set should then be performed. Work is in progress in this direction.\\
Concerning the $C_6$ coefficients, the differences between RSDHf and
MP2-srLDA values are -0.027, +0.55, +9.28 and +14.8 for He$_2$, Ne$_2$,
Ar$_2$ and Kr$_2$,
respectively. Comparison can be made with the difference between
the CCSD(T)-srDFT and MP2-srDFT values of Toulouse {\it et
al.}~\cite{PRA10_Julien_rpa-srDFT}: +0.49, +1.28,
+4.3 and +1.0, respectively. One could have expected these differences to be larger in
absolute value than the previous ones, as RSDHf
describes the long-range correlation within MP2, like in MP2-srDFT.
For Ne$_2$ and the heavier dimers, the lr-sr MP2 coupling term might not
be completely negligible at large distance, even though its contribution to the interaction energy is significantly smaller than
the purely long-range MP2 one. This should be investigated further on a
larger test set, including for example the benzene dimer. Replacing 
the long-range MP2 treatment with a long-range RPA one in the RSDHf model
would then be interesting to
investigate~\cite{JCP10_Wuming_rpa-srdft_weak_int}. Work is in progress in this
direction.
Let us finally mention that the advantages of MP2-srDFT with respect to
the BSSE and the basis set convergence are preserved in the RSDHf model as both
methods rely on the same wavefunction perturbation expansion. This
is illustrated in Fig.~\ref{bssecurve} for Ar$_2$, where the BSSE appears to be even
smaller at the RSDHf level than within MP2-srLDA, and in Fig.~\ref{bsconv} for
He$_2$ and Ne$_2$,
where the basis set convergence is shown to be much faster for both
RSDHf and MP2-srLDA than standard MP2 and CCSD(T).

\section{Conclusion}\label{Ccl}

The alternative decomposition of the short-range exchange-correlation
energy initially proposed by Toulouse {\it et
al.}~\cite{Toulouse2005TCA} has been used in the context of
range-separated density-functional perturbation theory. An exact energy
expression has been derived through second order and a connection with
conventional double hybrid density-functionals has been made. When
neglecting the second-order correction to the density, a new type of
range-separated double hybrid (RSDH) density-functional is obtained. It
is referred to as RSDHf where f stands for "full-range" as the regular
full-range interaction appears explicitly in the energy expression that is expanded
through second order. Its
specificity relies on (i) the use of an exact short-range
exchange energy, and (ii) the description of the coupling between long- and
short-range correlations at the MP2 level. Promising results were
obtained with the adapted LDA-type short-range correlation functional of
Paziani {\it et al.}~\cite{Paziani2006PRB} for the calculation of
interaction energies in the noble-gas dimers. RSDHf keeps all the
advantages of standard RSDH functionals, namely a small BSSE and a
faster convergence with respect to the basis set. The method can still be
improved in terms of accuracy. The first improvement could come from the orbitals
and their energies. In this respect it would be worth combining long-range HF with short-range
optimized effective potential approaches. A more flexible two-parameter extension, which makes a smooth connection
between RSDHf and conventional RSDH functionals, is also proposed and
should be tested on a larger test set.
Finally, replacing the long-range MP2 in RSDHf with a long-range RPA
description might be of interest for describing weakly interacting
stacked complexes such as the benzene dimer. Work is currently in
progress in those directions.

\subsection*{Acknowledgements}

The authors thank ANR (DYQUMA project). E.F. thanks Trygve Helgaker
and Andrew Teale for fruitful discussions. 

\appendix

\section{TWO-PARAMETER RSDHf MODEL}\label{2rsdhf_appendix}
A two parameter extension of the RSDHf model can be obtained when using
the following decomposition of the exact srHxc density-functional energy
\begin{eqnarray}\label{EsrHxc_decomps2_lambda}
\subsupi{E}{Hxc}{sr,\mu}[n]&=&
\lambda\langle\Psi^{\rm \mu}[n]|\subsupi{\hat{W}}{ee}{sr,\mu}|\Psi^{\rm \mu}[n]\rangle
+\subsupi{{E}}{\rm Hxc}{sr,\mu,\lambda}[n],
\end{eqnarray}
where $0\leq \lambda\leq 1$ and, according to Eq.~(\ref{EsrHxc_decomps2}), 
\begin{eqnarray}\label{EsrHxc_decomps2_lambda2}
\subsupi{E}{Hxc}{sr,\mu,\lambda}[n]&=&
(1-\lambda)\subsupi{{E}}{\rm Hxc}{sr,\mu}[n]+\lambda\subsupi{E}{\rm c,
md}{sr,\mu}[n]. 
\end{eqnarray}
With such a partitioning, the {\it exact} ground-state energy can be
rewritten as 
\begin{eqnarray}\label{rsHlambda}\begin{array}{l}
E=\langle\Psi^\mu|\hat{T}+
\subsupi{\hat{W}}{ee}{lr,\mu}+\lambda\subsupi{\hat{W}}{ee}{sr,\mu}
+\subi{\hat{V}}{ne}
|\Psi^\mu\rangle\\
\\
\hspace{0.7cm}+
(1-\lambda)\subsupi{{E}}{\rm Hxc}{sr,\mu}[n_{\Psi^\mu}]+\lambda\subsupi{E}{\rm c,
md}{sr,\mu}[n_{\Psi^\mu}]. 
\end{array}
\end{eqnarray}
From the Taylor expansion in $\alpha$ of the auxiliary energy
\begin{eqnarray}\label{newptalphalambda}\begin{array}{l}
\displaystyle 
\tilde{E}^{\alpha,\mu,\lambda}
=
{E}^{\alpha,\mu}
-\lambda\subsupi{E}{Hxc}{sr,\mu}[n_{\Psi^{\alpha,\mu}}]
\\
\\
\hspace{1.3cm}+\alpha\dfrac{\langle\Psi^{\alpha,\mu}|\lambda\subsupi{\hat{W}}{ee}{sr,\mu}|\Psi^{\alpha,\mu}\rangle}{\langle\Psi^{\alpha,\mu}|\Psi^{\alpha,\mu}\rangle}+\lambda\subsupi{E}{c,md}{sr,\mu}[n_{\Psi^{\alpha,\mu}}],
\end{array}
\end{eqnarray}
we obtain, through second order, a two-parameter RSDHf (2RSDHf) energy expression
\begin{eqnarray}\label{2rsdhfener}\begin{array}{l}
{E}_{\rm 2RSHf}
=\langle\Phi_0^\mu|\hat{T}+
\subsupi{\hat{W}}{ee}{lr,\mu}+\lambda\subsupi{\hat{W}}{ee}{sr,\mu}
+\subi{\hat{V}}{ne}
|\Phi_0^\mu\rangle
\\
\\
\hspace{1.5cm}+(1-\lambda)\subsupi{{E}}{\rm Hxc}{sr,\mu}[n_{\Phi_0^\mu}]+
\lambda\subsupi{E}{c,md}{sr,\mu}[n_{\Phi_0^\mu}]
\\
\\
\displaystyle
\hspace{1.5cm}+
\sum_{ij,ab}
\Bigg[\Big(V^{ij}_{ab}\Big)^{\rm lr,\mu}+2\lambda\Big(V^{ij}_{ab}\Big)^{\rm
sr,\mu}\Bigg]
\Big(t^{ab(1)}_{ij}\Big)^{\rm lr,\mu}
,
\end{array}
\end{eqnarray}
where the second-order corrections to the density have been neglected.
The corresponding approximate 2RSDHf exchange-correlation energy
expression is given in Eq.~(\ref{xcener_2rsdhf}).

\section{RANGE SEPARATION OF THE MP2 CORRELATION ENERGY FOR SMALL $\mu$ 
}\label{rsmp2_Taylor_mu_appendix}
In this appendix, a Taylor expansion of the purely long-range MP2
correlation energy and its coupling with the short-range correlation is
derived for small $\mu$ values. Note that HF-srDFT orbitals will be
denoted here with a superscript $"\mu"$. This will enable us to
distinguish them from the KS orbitals to which they reduce in the $\mu=0$ limit. 
According to Eq.~(\ref{0thord-SCeq}),  
the canonical doubly-occupied
$i^{\mu},j^{\mu}$ and unoccupied
$a^{\mu},b^{\mu}$ HF-srDFT orbitals fulfill the following long-range HF-type
equation
\begin{eqnarray}\label{hftypeeq}\begin{array} {l}
\left(\hat{t}+\hat{v}_{\rm ne}+\hat{v}_{\rm H}+\hat{u}^{\rm lr,\mu}_{\rm
HFX}+\hat{v}^{\rm sr,\mu}_{\rm xc}\right)\vert
p^{\mu}\rangle=\varepsilon^{\mu}_p\vert p^{\mu}\rangle,
\end{array}
\end{eqnarray}
where $\hat{v}_{\rm H}$, $\hat{u}^{\rm lr,\mu}_{\rm HFX}$ and $\hat{v}^{\rm sr,\mu}_{\rm
xc}$ correspond, respectively, to the local Hartree potential, the non-local long-range HF
exchange (HFX) potential and the local short-range exchange-correlation
potential. 
For analysis purposes, we expand the doubly-occupied orbitals and their
energies in perturbation
theory, using as unperturbed orbitals the KS doubly-occupied
$i$ and unoccupied $a$ orbitals, which are recovered in the $\mu=0$ limit: 
\begin{eqnarray}\label{pt12}\begin{array} {l}
\displaystyle\vert i^{\mu}\rangle=\vert
i\rangle+
\sum_a\vert a\rangle\frac{\langle a \vert\hat{u}^{\rm
lr,\mu}_{\rm HFX}+\hat{v}^{\rm sr,\mu}_{\rm xc}-\hat{v}_{\rm xc}
\vert i\rangle
}{\varepsilon_i-\varepsilon_a}
+\ldots,
\\
\\
\displaystyle\varepsilon^{\mu}_i=\varepsilon_i+\langle i \vert
\hat{u}^{\rm lr,\mu}_{\rm HFX}+\hat{v}^{\rm sr,\mu}_{\rm
xc}-\hat{v}_{\rm xc}\vert i
\rangle\\
\\
\displaystyle
\hspace{1.2cm}+\sum_a\frac{\langle a \vert\hat{u}^{\rm
lr,\mu}_{\rm HFX}+\hat{v}^{\rm sr,\mu}_{\rm xc}-\hat{v}_{\rm xc}
\vert i\rangle^2
}{\varepsilon_i-\varepsilon_a}+\ldots,
\\
\\
\displaystyle\langle p \vert \hat{u}^{\rm lr,\mu}_{\rm HFX}\vert q
\rangle=-\sum_j\langle pj \vert jq\rangle^{\rm lr,\mu},
\end{array}
\end{eqnarray}
where $\hat{v}_{\rm xc}$ is the standard exchange-correlation potential
operator. Note that, for simplicity, self-consistency in
Eq.~(\ref{hftypeeq}) is neglected. From the Taylor expansion about $\mu=0$ of the long-range interaction based on the error
function~\cite{Toulouse2004PRA}, 
\begin{eqnarray}\label{taylorexplr}\begin{array} {l}
\hspace{-0.5cm}\displaystyle 
w^{\rm lr,\mu}_{\rm ee}(r_{12})=
\frac{2}{\sqrt{\pi}}\Big(\mu-\frac{1}{3}\mu^3r_{12}^2+\frac{1}{10}\mu^5r_{12}^4+\mathcal{O}(\mu^7)\Big),\\
\end{array}
\end{eqnarray}
we obtain the following expression for the long-range HFX potential
matrix elements:
\begin{eqnarray}\label{taylorexplrHFX}\begin{array} {l}
\displaystyle\langle a\vert\hat{u}^{\rm lr,\mu}_{\rm HFX}\vert i\rangle
\displaystyle=\frac{2\mu^3}{3\sqrt{\pi}}\sum_j\langle ij \vert r_{12}^2 \vert
ja\rangle
\\
\\
\displaystyle \hspace{1.9cm}-\frac{\mu^5}{5\sqrt{\pi}}\sum_j\langle ij \vert r_{12}^4
\vert ja\rangle +
\mathcal{O}(\mu^7)
. 
\displaystyle 
\end{array}
\end{eqnarray}
Moreover, since the short-range exchange and correlation
density-functional energies can be
expanded as~\cite{Toulouse2004PRA}
\begin{eqnarray}\label{srXCtalorexp}\begin{array} {l}
\displaystyle 
E^{\rm sr,\mu}_{\rm x}=E_{\rm x}+\frac{\mu N}{\sqrt{\pi}}-E^{(3)\rm
lr}_{\rm x}\mu^3-E^{(5)\rm lr}_{\rm x}\mu^5+\mathcal{O}(\mu^7),
\\
\\
E^{\rm sr,\mu}_{\rm c}=E_{\rm c}-E^{(6)\rm lr}_{\rm
c}\mu^6+\mathcal{O}(\mu^7),
\end{array}
\end{eqnarray}
where $N$ denotes the number of electrons,
the potential operator difference $\hat{v}^{\rm sr,\mu}_{\rm
xc}-\hat{v}_{\rm xc}$ can be rewritten as 
\begin{eqnarray}\label{diffvxc}\begin{array} {l}
\displaystyle 
\hat{v}^{\rm sr,\mu}_{\rm xc}-\hat{v}_{\rm
xc}=\frac{\mu}{\sqrt{\pi}}-\hat{v}^{(3)\rm
lr}_{\rm x}\mu^3-\hat{v}^{(5)\rm lr}_{\rm
x}\mu^5\\
\\
\hspace{2cm}-\hat{v}^{(6)\rm lr}_{\rm c}\mu^6 +\mathcal{O}(\mu^7).
\end{array}
\end{eqnarray}
From Eqs.~(\ref{pt12}), (\ref{taylorexplrHFX}) and (\ref{diffvxc}),
we conclude that
\begin{eqnarray}\label{orbentaylorexp1}\begin{array} {l}
\displaystyle 
\vert i^{\mu}\rangle=\vert i\rangle+\mathcal{O}(\mu^3),
\end{array}
\end{eqnarray}
and
\begin{eqnarray}\label{orbentaylorexp2}\begin{array} {l}
 \displaystyle\varepsilon^{\mu}_i=\varepsilon_i-\frac{\mu}{\sqrt{\pi}}
+\mu^3\varepsilon^{(3)}_i
 +\mathcal{O}(\mu^4),\\
 \\\displaystyle 
\varepsilon^{(3)}_i=
-\langle
 i\vert\hat{v}_{\rm x}^
 {(3)\rm lr}\vert
 i\rangle+\frac{2}{3\sqrt{\pi}}\sum_j\langle ij\vert r_{12}^2 \vert
 ji\rangle.
\end{array}
\end{eqnarray}

According to Eqs.~(\ref{taylorexplr}) and (\ref{orbentaylorexp1}), the 
long-range integrals squared computed with the HF-srDFT orbitals are
therefore expanded as 
\begin{eqnarray}\label{lrlrint}\begin{array} {l}
\displaystyle
\Big(\langle {a}^{\mu}{b}^{\mu} \vert
i^{\mu}j^{\mu}\rangle^{\rm lr,\mu}\Big)^2=
\displaystyle\frac{4\mu^6}{9\pi}\langle ab \vert
r_{12}^2 \vert ij\rangle^2
+\mathcal{O}(\mu^7),
\end{array}
\end{eqnarray}
and the long-/short-range integrals product equals 
\begin{eqnarray}\label{lrsrint}\begin{array} {l}
\displaystyle
\langle {a}^{\mu}{b}^{\mu} \vert i^{\mu}j^{\mu}\rangle^{\rm lr,\mu}
\langle {a}^{\mu}{b}^{\mu} \vert i^{\mu}j^{\mu}\rangle^{\rm sr,\mu}
=\\
\\\displaystyle
-\frac{2\mu^3}{3\sqrt{\pi}}\langle ab \vert r_{12}^2 \vert
ij\rangle\langle ab\vert ij\rangle
+\\
\\
\displaystyle\frac{\mu^5}{5\sqrt{\pi}}\langle ab \vert r_{12}^4 \vert
ij\rangle\langle ab\vert ij\rangle
\\
\\
-
\displaystyle\frac{4\mu^6}{9\pi}\langle ab \vert
r_{12}^2 \vert ij\rangle^2
+\mathcal{O}(\mu^7).
\end{array}
\end{eqnarray}
Moreover, according to Eq.~(\ref{orbentaylorexp2}), the HF-srDFT orbital energy
differences can be rewritten as
\begin{eqnarray}\label{delataepsilontaylorexp}
\hspace{-1cm}{\Delta\varepsilon^{ab,\mu}_{ij}}&=&
\varepsilon^{\mu}_i+\varepsilon^{\mu}_j-\varepsilon^{\mu}_a-\varepsilon^{\mu}_b
\nonumber\\
&=&{\Delta\varepsilon^{ab}_{ij}}-\frac{2\mu}{\sqrt{\pi}}
+\mu^3\left(\varepsilon^{(3)}_i+\varepsilon^{(3)}_j\right)
 +\mathcal{O}(\mu^4),
\end{eqnarray}
so that, from the Taylor expansion
\begin{eqnarray}\label{inversedelataepsilon}\begin{array} {l}
\displaystyle
\frac{1}{\Delta\varepsilon^{ab,\mu}_{ij}}=
\frac{1}{\Delta\varepsilon^{ab}_{ij}}
\left(
1-
\frac{\Delta\varepsilon^{ab,\mu}_{ij}-\Delta\varepsilon^{ab}_{ij}}{\Delta\varepsilon^{ab}_{ij}}
\right.\\
\\
\displaystyle
\hspace{2.9cm}+\left.\left(
\frac{\Delta\varepsilon^{ab,\mu}_{ij}-\Delta\varepsilon^{ab}_{ij}}{\Delta\varepsilon^{ab}_{ij}}
\right)^2+\ldots
\right),
\end{array}
\end{eqnarray}
their inverse can be expanded as
\begin{eqnarray}\label{inversedelataepsilon2}\begin{array} {l}
\displaystyle
\frac{1}{\Delta\varepsilon^{ab,\mu}_{ij}}=
\frac{1}{\Delta\varepsilon^{ab}_{ij}}+
\frac{2\mu}{\left(\Delta\varepsilon^{ab}_{ij}\right)^2\sqrt{\pi}}
+\frac{4\mu^2}{\left(\Delta\varepsilon^{ab}_{ij}\right)^3{\pi}}
\\
\\
\displaystyle
\hspace{2.5cm}-\frac{\mu^3\left(\varepsilon^{(3)}_i+\varepsilon^{(3)}_j\right)}{\left(\Delta\varepsilon^{ab}_{ij}\right)^2}+\mathcal{O}(\mu^4).
\end{array}
\end{eqnarray}
Using Eqs.~(\ref{lrlrint}), (\ref{lrsrint}) and
(\ref{inversedelataepsilon2}), we finally obtain the expansions in
Eq.~(\ref{rsmp2correnerexpansion})
where the Taylor expansion coefficients are expressed in terms of the KS
orbitals and orbital energies as follows:
\begin{eqnarray}\label{rsmp2correnerexpansionAB}\begin{array} {l}
A_6=\displaystyle\frac{4}{9\pi}
\displaystyle\sum_{a,b,i,j}\frac{\langle ab \vert r_{12}^2 \vert
ij\rangle
}{
\Delta\varepsilon^{ab}_{ij}
}
\left(2\langle ab
\vert r_{12}^2 \vert ij\rangle-\langle ab \vert r_{12}^2 \vert
ji\rangle\right),
\\
\\
B_3=-
\displaystyle\frac{4}{3\sqrt{\pi}}
\displaystyle\sum_{a,b,i,j}\frac{\langle ab \vert r_{12}^2 \vert
ij\rangle
}
{\Delta\varepsilon^{ab}_{ij}}
\left(2\langle ab
\vert ij\rangle-\langle ab \vert 
ji\rangle\right),
\\
\\
B_4=-
\displaystyle\frac{8}{3{\pi}}
\displaystyle\sum_{a,b,i,j}\frac{\langle ab \vert r_{12}^2 \vert
ij\rangle
}
{\left(\Delta\varepsilon^{ab}_{ij}\right)^2}
\left(2\langle ab
\vert ij\rangle-\langle ab \vert 
ji\rangle\right),
\\
\\
B_5=
\displaystyle\frac{2}{5\sqrt{\pi}}
\displaystyle\sum_{a,b,i,j}\frac{\langle ab \vert r_{12}^4 \vert
ij\rangle
}
{\Delta\varepsilon^{ab}_{ij}}
\left(2\langle ab
\vert ij\rangle-\langle ab \vert 
ji\rangle\right)\\
\\
\hspace{0.8cm}-\displaystyle\frac{16}{3{\pi}\sqrt{\pi}
}
\displaystyle\sum_{a,b,i,j}\frac{\langle ab \vert r_{12}^2 \vert
ij\rangle
}
{\left(\Delta\varepsilon^{ab}_{ij}\right)^3}
\left(2\langle ab
\vert ij\rangle-\langle ab \vert 
ji\rangle\right),
\\
\\
B_6=-2A_6
+\displaystyle\frac{4}{5{\pi}}
\displaystyle\sum_{a,b,i,j}\frac{\langle ab \vert r_{12}^4 \vert
ij\rangle
}
{\left(\Delta\varepsilon^{ab}_{ij}\right)^2}
\left(2\langle ab
\vert ij\rangle-\langle ab \vert 
ji\rangle\right)\\
\\
+\displaystyle\frac{4}{3\sqrt{\pi}}
\displaystyle\sum_{a,b,i,j}\frac{\left(\varepsilon^{(3)}_i+\varepsilon^{(3)}_j\right)\langle ab \vert r_{12}^2 \vert
ij\rangle
}
{\left(\Delta\varepsilon^{ab}_{ij}\right)^2}
\left(2\langle ab
\vert ij\rangle-\langle ab \vert 
ji\rangle\right).
\end{array}
\end{eqnarray}



\newcommand{\Aa}[0]{Aa}
%


\clearpage

\textbf{FIGURE AND TABLE CAPTIONS}

\begin{figure}
\caption{\label{Emu_atcurves} Cornaton {\it et al.}, Physical Review A}
\hspace{1cm}
\vspace{-4cm}
\begin{center}
\begin{tabular}{c}
\resizebox{18cm}{!}{\includegraphics{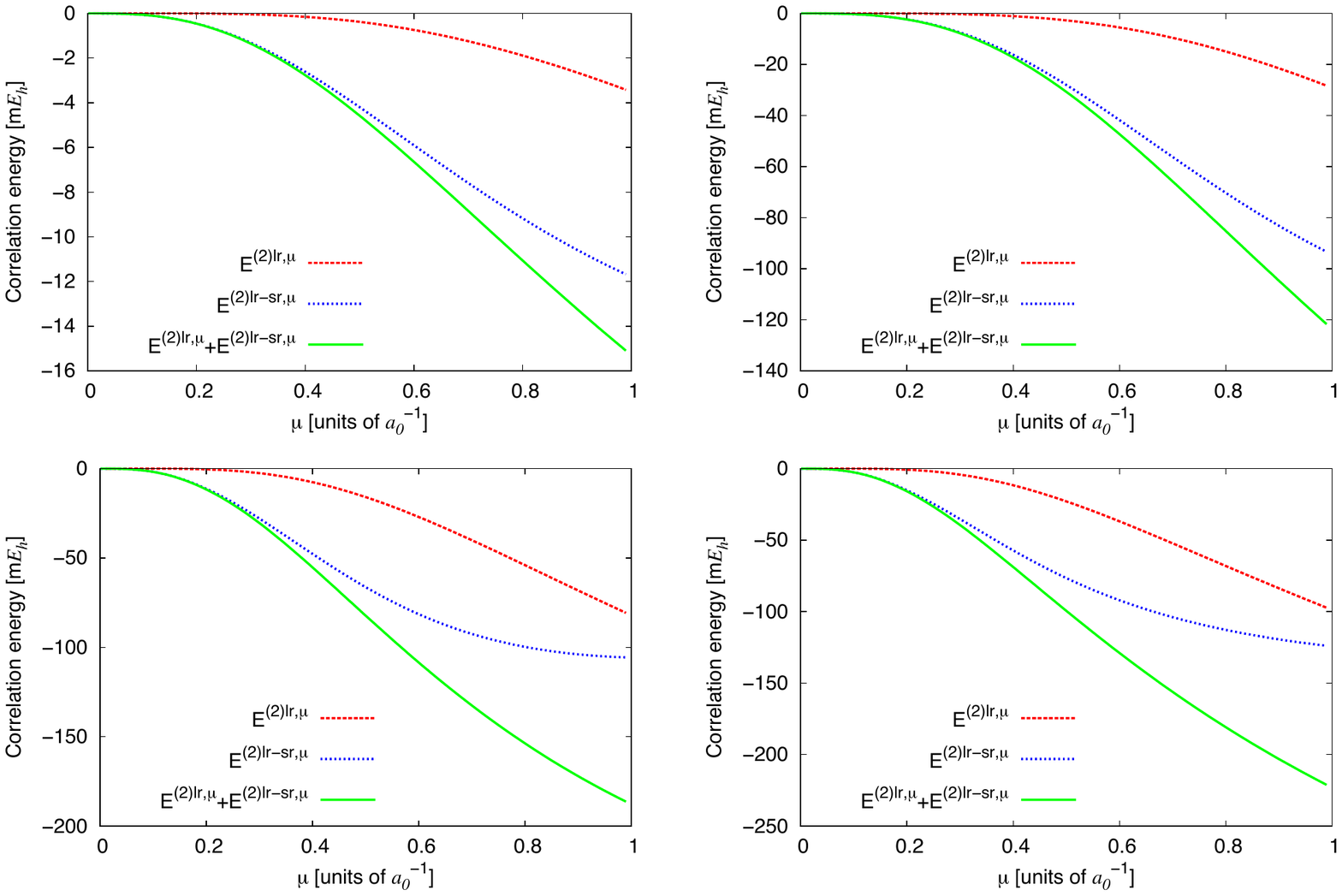}}
\end{tabular}
\end{center}
  \end{figure}

\begin{figure}
 \caption{\label{potcurve-comp} Cornaton {\it et al.}, Physical Review A}
\vspace{-4cm}
\begin{center}
\begin{tabular}{c}
\hspace{-1.0cm}
\resizebox{19cm}{!}{\includegraphics{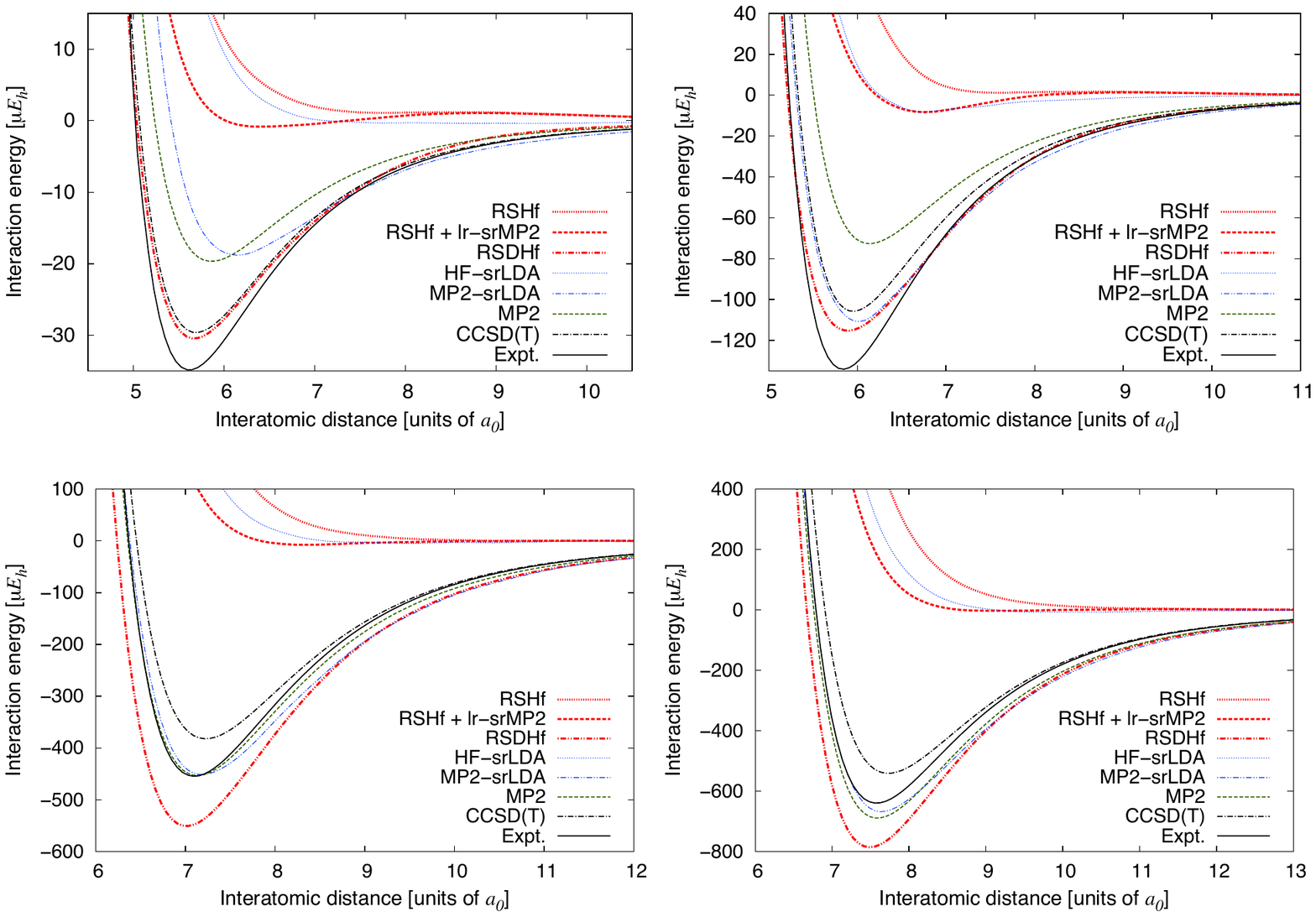}}
\end{tabular}
\end{center}
\end{figure}

\begin{figure}
\caption{\label{Emu_Ar2curves} Cornaton {\it et al.}, Physical Review A}
\vspace{-6cm}
\begin{center}
\begin{tabular}{c}
\hspace{-1.5cm}\resizebox{19cm}{!}{\includegraphics{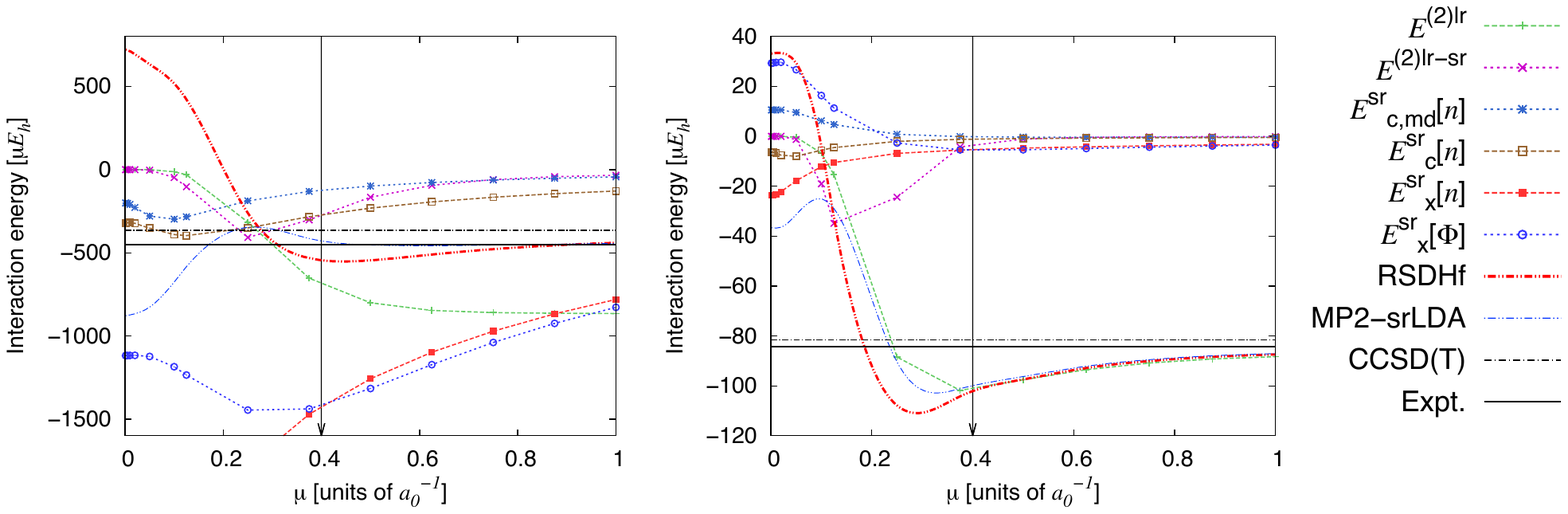}}
\end{tabular}
\end{center}
  \end{figure}

\begin{figure}
\begin{center}
\caption{\label{bssecurve}  Cornaton {\it et al.}, Physical Review A}
\subfloat{\label{Ar2-bsse}\includegraphics[width=0.8\textwidth]{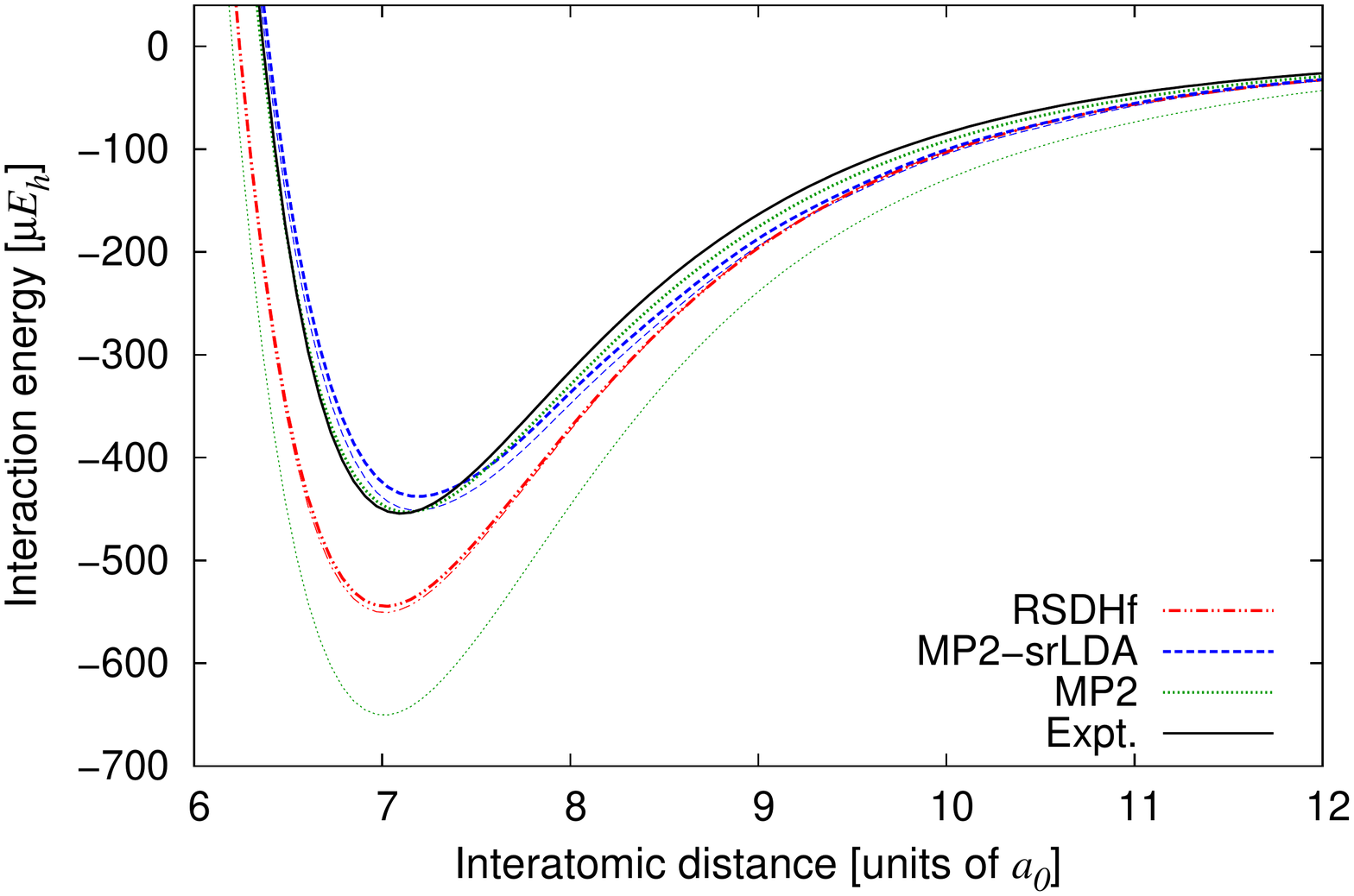}}
\end{center}
\end{figure}

\begin{figure}
\caption{\label{bsconv} Cornaton {\it et al.}, Physical Review A}
\vspace{-6cm}
\begin{center}
\begin{tabular}{c}
\hspace{-1.5cm}\resizebox{19cm}{!}{\includegraphics{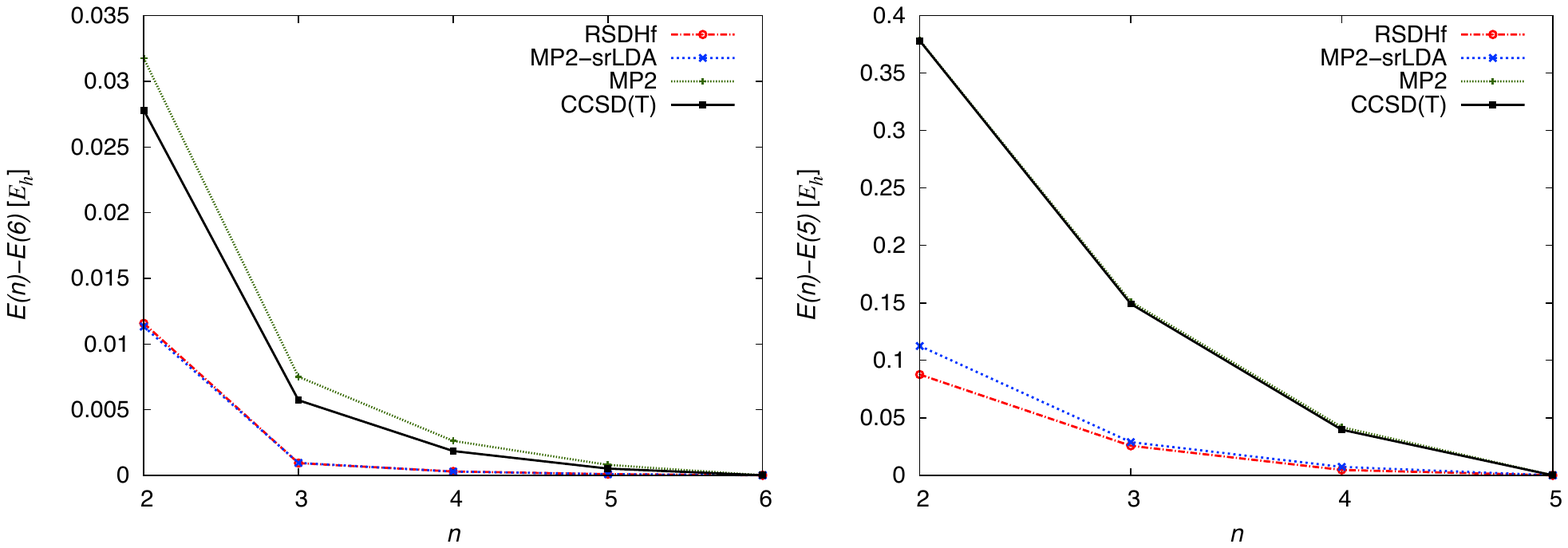}}
\end{tabular}
\end{center}
\end{figure}

\begin{description}
 \item[Figure \ref{Emu_atcurves}] (Color online) Long-range $E^{(2)\rm
 lr,\mu}$ (dashed red line) and
 long-/short-range $E^{(2)\rm
lr-sr,\mu}$ (dotted blue line) MP2 correlation energies, as well as the
sum of the two contributions (solid green line), computed with the HF-srDFT
orbitals
and orbital energies for He (top left), Ne (top right), Ar (bottom left)
and Kr (bottom right) atoms when varying the $\mu$ parameter.

 \item[Figure \ref{potcurve-comp}] (Color online) Interaction energy curves obtained
 for He$_2$ (top left), Ne$_2$ (top right), Ar$_2$ (bottom left) and Kr$_2$ (bottom right) with conventional srLDA schemes (thin dotted and double-dotted-dashed blue lines) and the new range-separated models (thick red lines). See text for further details. Comparison is made with MP2 (thin dashed green line), CCSD(T) (dotted-dashed black line) and the
 experimental (solid black line) results of Ref.~\cite{Tang2003JCP}. The $\mu$ parameter is set to $0.4~a_0^{-1}$.

 \item[Figure \ref{Emu_Ar2curves}] (Color online) Decomposition of the 
 $\mu$-dependent RSDHf (thick double-dotted-dashed red line) and
 MP2-srLDA (thin double-dotted-dashed blue line) interaction energies for Ar$_2$ at both $R=7.013$ (left)
 and $10.0a_0$ (right) bond distances. $E^{(2)\rm lr}$, $E^{(2)\rm lr-sr}$,
 $E_{\rm x}^{\rm sr}[\Phi]=
-\sum_{ij}\langle ij \vert ji\rangle^{\rm sr,\mu}$ and $n$ denote the
purely long-range MP2 interaction energy, the long-/short-range coupling
contribution, the exact short-range exchange energy contribution
computed with the HF-srDFT orbitals, and
the HF-srDFT density, respectively. See text for further details.
Comparison is made with conventional CCSD(T) (dotted-dashed black line)
and experiment (black solid line). 
 

 \item[Figure \ref{bssecurve}] (Color online) RSDHf (red double-dotted-dashed lines), MP2-srLDA (blue dashed lines)
 and regular MP2 (green dotted lines) interaction energy curves of Ar$_2$ with (thick lines) and without (thin lines) BSSE correction. The $\mu$ parameter is set to $0.4~a_0^{-1}$. 

 \item[Figure \ref{bsconv}] (Color online) Basis set (aug-cc-pV$n$Z) convergence for the RSDHf and
 MP2-srLDA total energies in He$_2$ (left) and Ne$_2$ (right). Experimental equilibrium
 distances~\cite{Tang2003JCP} and $\mu=0.4~a_0^{-1}$ were used.
 Comparison is made with conventional MP2 and CCSD(T).

\end{description}

\clearpage

\begin{description}


 \item[Table \ref{tabcurvref}] Equilibrium bond distances ($R_{\rm
 e}/a_0$), binding energies ($D_{\rm e}/\mu E_{\rm h}$), harmonic
 vibrational wavenumbers ($\omega_{\rm e}/{\rm cm}^{-1}$), $C_6$
 dispersion coefficients ($C_6/a_0^6$) and hard core radii
 ($\sigma/a_0$) computed for the first four noble-gas dimers at the
 RSDHf and MP2-srLDA levels. Comparison is made with MP2 and CCSD(T)
 values. Reduced constants, which are obtained when dividing by the accurate
 "experimental"
 values of Ref.~\cite{Tang2003JCP}, are given in parentheses.

\end{description}
\clearpage

\clearpage

  

\clearpage

%

\clearpage

\clearpage

\clearpage





\clearpage

\clearpage

\begin{table}
\begin{center}
\caption{\label{tabcurvref} Cornaton {\it et al.}, Physical Review A}
\begin{tabular}{rc|*{5}{c}}
\hline
\hline
&&MP2&CCSD(T)&MP2-srLDA&RSDHf&Expt.$^a$\\
\hline
\multirow{5}{*}{He$_2$}&$R_{\rm e}$&5.862 (1.043)&5.690 (1.013)&6.147 (1.094)&5.670 (1.009)&5.618 (1.000)\\
&$D_{\rm e}$&19.67 (0.564)&29.63 (0.850)&18.81 (0.539)&30.47 (0.874)&34.87 (1.000)\\
&$\omega_{\rm e}$&24.9 (0.780)&30.3 (0.886)&21.2 (0.620)&31.1 (0.909)&34.2 (1.000)\\
&$C_6$&1.203 (0.815)&1.467 (0.994)&1.613 (1.093)&1.586 (1.075)&1.476 (1.000)\\
&$\sigma$&5.235 (1.043)&5.066 (1.010)&5.414 (1.079)&5.037 (1.004)&5.018 (1.000)\\
\hline
\hline
\multirow{5}{*}{Ne$_2$}&$R_{\rm e}$&6.138 (1.051)&5.952 (1.019)&6.015 (1.030)&5.893 (1.009)&5.839 (1.000)\\
&$D_{\rm e}$&72.70 (0.542)&105.83 (0.789)&110.87 (0.826)&115.29 (0.859)&134.18 (1.000)\\
&$\omega_{\rm e}$&20.7 (0.704)&25.7 (0.874)&24.3 (0.827)&22.3 (0.758)&29.4 (1.000)\\
&$C_6$&5.320 (0.821)&7.967 (1.229)&6.819 (1.052)&7.369 (1.137)&6.480 (1.000)\\
&$\sigma$&5.505 (1.053)&5.331 (1.019)&5.310 (1.015)&5.213 (0.997)&5.230 (1.000)\\
\hline
\hline
\multirow{5}{*}{Ar$_2$}&$R_{\rm e}$&7.133 (1.005)&7.227 (1.018)&7.183 (1.012)&7.013 (0.988)&7.099 (1.000)\\
&$D_{\rm e}$&453.08 (0.997)&382.13 (0.841)&451.32 (0.993)&550.71 (1.212)&454.50 (1.000)\\
&$\omega_{\rm e}$&30.5 (0.953)&27.9 (0.872)&28.6 (0.893)&33.8 (1.056)&32.0 (1.000)\\
&$C_6$&77.56 (1.174)&66.10 (1.001)&80.43 (1.217)&89.71 (1.358)&66.07 (1.000)\\
&$\sigma$&6.357 (0.998)&6.454 (1.014)&6.381 (1.002)&6.237 (0.980)&6.367 (1.000)\\
\hline
\hline
\multirow{4}{*}{Kr$_2$}&$R_{\rm e}$&7.587 (1.001)&7.729 (1.020)&7.642 (1.008)&7.491 (0.989)&7.578 (1.000)\\
&$D_{\rm e}$&689.18 (1.078)&541.08 (0.846)&668.02 (1.045)&785.60 (1.229)&639.42 (1.000)\\
&$\omega_{\rm e}$&24.3 (0.996)&21.6 (0.889)&24.2 (0.996)&26.2 (1.078)&24.3 (1.000)\\
&$C_6$&165.09 (1.230)&129.26 (0.963)&160.30 (1.195)&175.07 (1.305)&134.19 (1.000)\\
&$\sigma$&6.760 (0.996)&6.906 (1.017)&6.785 (0.999)&6.668 (0.982)&6.789 (1.000)\\
\hline
\hline
\end{tabular}\newline
$^a$ Reference~\cite{Tang2003JCP}
\end{center}
\end{table}

\end{document}